%% file: root.tex
\documentclass[letterpaper, 10 pt, conference]{ieeeconf}  % Comment this line out if you need a4paper
\usepackage{cite}
\usepackage{amsmath,amssymb,amsfonts}
\usepackage{algorithmic}
\usepackage{graphicx}
\usepackage{algorithm}
\usepackage{subcaption}
\usepackage{siunitx}
\usepackage{multirow}
\usepackage{mathtools}

\usepackage{tkz-euclide}
\usetikzlibrary{arrows}

\usepackage{bbm}
\usepackage[hidelinks]{hyperref}
\usepackage{siunitx}
\usepackage{adjustbox}

\newcommand{\xx}{\mathbf{x}}
\newcommand{\bxx}{\Bar{\mathbf{x}}}

\newcommand{\zz}{\mathbf{z}}

\newcommand{\uu}{\mathbf{u}}
\newcommand{\buu}{\Bar{\mathbf{u}}}

\newcommand{\OO}{\mathbb{O}}
\newcommand{\VV}{\mathbb{V}}
\newcommand{\RR}{\mathbb{R}}
\newcommand{\NN}{\mathbb{N}}

\newcommand{\tx}{\mathrm}

\newcommand{\pp}{\mathbf{p}}

\newcommand\rsmraise[1]{%
  \ifx#1\displaystyle .8\else
    \ifx#1\textstyle .8\else
      \ifx#1\scriptstyle .6\else
        .45%
      \fi
    \fi
  \fi}

  \DeclareMathSymbol{\shortminus}{\mathbin}{AMSa}{"39}

\newcommand{\nquad}{\! \! \! \! \! \! \! \!}
\newcommand{\ncol}{\! \!}

\newtheorem{remark}{Remark}
\AfterEndEnvironment{Remark}{\noindent\ignorespaces}
\newtheorem{proposition}{Proposition}
\AfterEndEnvironment{Proposition}{\noindent\ignorespaces}

\AfterEndEnvironment{Lemma}{\noindent\ignorespaces}
\newtheorem{corollary}{Corollary}
\AfterEndEnvironment{Corollary}{\noindent\ignorespaces}

\newcolumntype{?}{!{\vrule width 1pt}}

\usepackage{xargs}                      % Use more than one optional parameter in a new commands
\usepackage[colorinlistoftodos,prependcaption,textsize=tiny]{todonotes}
\newcommandx{\unsure}[2][1=]{\todo[linecolor=red,backgroundcolor=red!25,bordercolor=red,#1]{#2}}
\newcommandx{\change}[2][1=]{\todo[linecolor=blue,backgroundcolor=blue!25,bordercolor=blue,#1]{#2}}
\newcommandx{\info}[2][1=]{\todo[linecolor=OliveGreen,backgroundcolor=OliveGreen!25,bordercolor=OliveGreen,#1]{#2}}
\newcommandx{\improvement}[2][1=]{\todo[linecolor=Plum,backgroundcolor=Plum!25,bordercolor=Plum,#1]{#2}}
\newcommandx{\thiswillnotshow}[2][1=]{\todo[disable,#1]{#2}}
\begin{document}

\title{Tight Collision Avoidance for Stochastic Optimal Control: with Applications in Learning-based, Interactive Motion Planning}

\author{Erik Börve, Nikolce Murgovski, and Leo Laine% <-this % stops a space
\thanks{This work was partially supported by the Wallenberg AI, Autonomous Systems and Software Program (WASP) funded by the Knut and Alice Wallenberg Foundation.}% <-this % stops a space
\thanks{Erik Börve and Nikolce Murgovski are with the Department of Electrical Engineering, Chalmers University of Technology, 412 96 Göteborg, Sweden {\tt\footnotesize\{borerik, nikolce.murgovski\}@chalmers.se}}%
\thanks{Leo Laine is with the Department of Mechanics and Maritime Science, Chalmers University of Technology, 412 96 Göteborg, Sweden {\tt\footnotesize leo.laine@chalmers.se}}%
\thanks{Erik Börve and Leo Laine is with the Department of Safe, and Efficient Driving, Volvo Group Trucks Technology, 417 10 Göteborg, Sweden {\tt\footnotesize \{erik.borve, leo.laine\}@chalmers.se}}%
}

% \markboth{IEEE TRANSACTIONS ON INTELLIGENT VEHICLES,~Vol.~XX, No.~X, Xxx~XXXX}%
% {Börve \MakeLowercase{\textit{et al.}}: Tight Collision Avoidance for Stochastic Optimal Control: with
% Applications in Learning-based, Interactive Motion Planning}

% \IEEEpubid{0000--0000/00\$00.00~\copyright~XXXX IEEE}

\maketitle

%%%%%%%%%%%%%%%%%%%%%%%%%%%%%%%%%%%%%%%%%%%%%%%%%%%%%%%%%%%%%%%%%%%%%%%%%%%%%%%%
\begin{abstract}
Trajectory planning in dense, interactive traffic scenarios presents significant challenges for autonomous vehicles, primarily due to the uncertainty of human driver behavior and the non-convex nature of collision avoidance constraints. This paper introduces a stochastic optimal control framework to address these issues simultaneously, without excessively conservative approximations. We opt to model human driver decisions as a Markov Decision Process and propose a method for handling collision avoidance between non-convex vehicle shapes by imposing a positive distance constraint between compact sets. In this framework, we investigate three alternative chance constraint formulations. To ensure computational tractability, we introduce tight, continuously differentiable reformulations of both the non-convex distance constraints and the chance constraints. The efficacy of our approach is demonstrated through simulation studies of two challenging interactive scenarios: an unregulated intersection crossing and a highway lane change in dense traffic. Supplementary animations are available at \cite{github}.
\end{abstract}

% \begin{IEEEkeywords}
% Autonomous driving, Optimization and control, Data-based approaches, Markov Decision Processes
% \end{IEEEkeywords}

% ---------------------------------------------------------------------------------------------
% ---------------------------------------------------------------------------------------------
% ---------------------------------------------------------------------------------------------
% ---------------------------------------------------------------------------------------------
% ---------------------------------------------------------------------------------------------

\section{Introduction} \label{sec:introduction}
Trajectory planning problems appear in a wide array of promising robotic applications, e.g., manipulators and UAVs \cite{blackmore2006optimal, borrelli2004collision}. In recent years, these methods have also been popularized for autonomous vehicles (AVs). For AVs, the primary control objective is to navigate while avoiding collisions with different obstacles, e.g., static objects, pedestrians, or other vehicles. A secondary objective is often to find a trajectory that is optimal with respect to some efficiency measure, e.g., travel time, passenger comfort, and energy consumption. A large body of work has shown that optimization-based control is an attractive solution that simultaneously tackles both these objectives; see, e.g., \cite{falcone2007predictive, liniger2015optimization, murgovski2015predictive} for applications.

\begin{figure}
    \centering
    \includegraphics[width = 0.5 \textwidth,trim={2cm 0 0 0},clip]{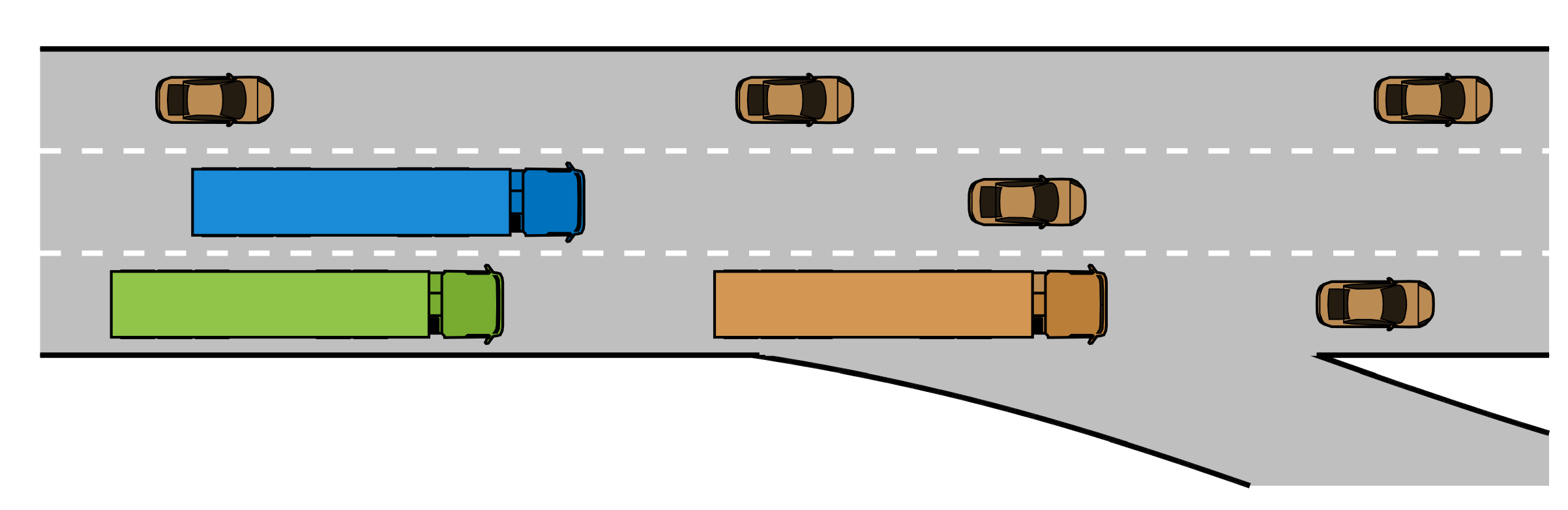}
    \caption{Forced lane-change scenario for an autonomous heavy-vehicle (blue) in dense traffic. To reach the upcoming exit ramp, the AV needs to interact with an adjacent heavy-vehicle (green).}
    \label{fig:exScenario}
\end{figure}

The deployment of such trajectory planners in real-world traffic scenarios presents a multitude of different challenges. In this paper, we mainly focus on two difficult problems. First, collision avoidance constraints are often non-convex, or even non-smooth, posing computational challenges for numerical optimization solvers. To meet real-time demands, it is crucial to formulate these constraints in a computationally efficient way \cite{zhang2020optimization}. Often, practitioners resort to convex, but conservative, outer approximations of the vehicles. Second, dynamic traffic environments introduce many sources of uncertainty, e.g., sensing and perception \cite{yang2023uncertainties}. Among the most challenging to capture, and the focus of this work, is the behavior of other human drivers. For example, choosing when to yield for a merging vehicle or when to change lane, can vary drastically for different human drivers in different scenarios \cite{wang2020non}. Fig. \ref{fig:exScenario} shows such a scenario, in which an autonomous heavy-vehicle combination (HVC) must negotiate with human drivers to successfully reach the exit of the highway. In such scenarios, human driver decisions are intrinsically dependent on how the AV interacts with them. Therefore, a safe and efficient AV must predict the behavior of other drivers, while also considering how the AV itself influences their behavior. As the decisions of other drivers may depend on countless factors, we wish to design a motion planner that explicitly treats uncertainties in human decision-making, while not imposing excessively conservative limitations on performance. To this end, learning-based stochastic optimal control with chance constraints has been identified as an attractive solution \cite{hewing2020learning}.

% ---------------------------------------------------------------------------------------------
% ---------------------------------------------------------------------------------------------

\subsection{Related Work}
A large amount of research has been dedicated to the trajectory planning problem, and many approaches exist. In this work, we focus solely on optimization-based approaches and refer the interested reader to \cite{lavalle2006planning} for more details on alternatives. In general, optimization-based approaches formulate the trajectory planning problem as an optimal control problem (OCP) over a finite time horizon, that aims to minimize a cost function subject to constraints, e.g., ego-vehicle dynamics, actuator limits, and collision avoidance. 

In environments with uncertain dynamic obstacles, their future states are predicted over the horizon and are, in turn, incorporated in the collision avoidance constraints.

\subsubsection{Environment Uncertainty}
 With vehicles operated by human drivers, their behavior may cause, e.g., their position and orientation to be inherently uncertain. Predicting the behavior of general uncontrollable agents is a notoriously difficult and well-explored area of research \cite{lefevre2014survey}. Recently, large-scale machine learning methods have gathered attention for obtaining state-of-the-art prediction accuracy on real-world data sets; see, e.g., \cite{trajectron++,ngiam2021scene}. However, directly incorporating these approaches into gradient-based numerical optimization solvers, a critical step for producing interactive motion plans, remains difficult \cite{borve2023interaction}. A promising alternative is to model the dynamics of other drivers with Markov Models \cite{schuurmans2023safe}. This approach assumes that drivers pick from an apriori known finite set of decisions, according to an apriori unknown distribution. A compelling property of this approach is that the multimodal nature observed in human decision-making is preserved \cite{chen2022interactive}. Importantly, by limiting the estimation to the decision distribution, the problem can be cast as a tractable chance-constrained stochastic OCP (S-OCP). More recent work, \cite{wang2023interaction}, further extends this approach by learning a decision distribution that considers vehicle interactions, obtaining an interactive motion planner.

\subsubsection{Collision Avoidance} The vast majority of existing work considers point-mass models and accounts for the vehicle shape with conservative expansions of constraints. Examples include convex inner approximations \cite{schoels2020nmpc}, constraining the radial distance between multiple ellipsoids \cite{schwarting2017parallel}, and non-linear constraint smoothing \cite{borve2023interaction}. However, these approaches are less applicable in scenarios with small collision margins, e.g., parking or dense traffic. In \cite{zhang2020optimization}, the authors address this problem by deriving constraints for convex polytopic objects by applying Lagrange duality theory on the distance function. The computational complexity is further reduced in \cite{dietz2023efficient,fan2023efficient} by considering vertex representations of the convex polytopes ($\mathcal{V}$-polytopes). 

\subsubsection{Chance Constraints}
The future states of the dynamic obstacles are uncertain, and we wish to obtain appropriate guarantees on collision avoidance. Enforcing the constraints robustly, i.e., accounting for any possible obstacle state, may give strong guarantees, but often come with large limitations on performance. Chance constraints are a popular alternative that allow constraint violations in unlikely scenarios, consequently imposing less performance limitations \cite{hewing2019cautious}. Although tractable chance constraint reformulations for all types of distributions do not exist, approximations are available in specific cases, e.g., distributions over discrete finite sets \cite{nemirovski2012safe}. Further, one may formulate chance constraints differently depending on the desired guarantees, e.g, over each step in the horizon \cite{blackmore2011chance} or jointly over the entire horizon \cite{ono2015chance}. The choice of chance constraint formulation should balance safety requirements with performance demands, but is rarely investigated thoroughly in prior work.

% ---------------------------------------------------------------------------------------------
% ---------------------------------------------------------------------------------------------

\subsection{Outline and Contribution}
As in prior work, we opt to model the human drivers as Markov Decision Processes with a finite set of possible interactive decisions. Unlike previous work, we investigate dense and slow-moving traffic scenarios, where an AV needs to interact with another human driver to successfully complete a maneuver. Further, we consider a setting where each vehicle may be described by non-convex sets, e.g., in the case of HVCs. To handle collision avoidance, we propose a method that can impose a positive distance constraint between any two closed and compact sets, represented in any spatial dimension. In this setting, we investigate three alternative chance constraint formulations for distributions on discrete finite sets and propose tight outer approximations for each of the alternatives. Importantly, the chance-constraint reformulations are independent of the complexity of the vehicle representations, improving the computational complexity. Finally, we investigate the efficacy of our methods by performing exhaustive simulations of two interactive scenarios with small collision margins: 1.) An unregulated intersection crossing; 2.) A complex lane change in dense highway traffic.

% ---------------------------------------------------------------------------------------------
% ---------------------------------------------------------------------------------------------
% ---------------------------------------------------------------------------------------------
% ---------------------------------------------------------------------------------------------
% ---------------------------------------------------------------------------------------------

\section{Problem Formulation} \label{sec:prob_formulation}
In the following section, we first present a general formulation for the interactive motion planing problem with MDPs as an intractable, discrete-time S-OCP. In later sections, we provide efficient reformulations, such that the problem becomes tractable. We consider $N_V$ human vehicles referred to with index $i \in \VV$ where $N_V = |\VV|$ and reserve the subscript $\tx{e}$ for the ego-vehicle. In particular, we focus on tractor-trailer vehicles as a representative type of HVCs. This study is limited to straight roads. An extension to curved roads is conceptually simple by representing the vehicle dynamics in the Frenet frame \cite{liniger2015optimization}.

% ---------------------------------------------------------------------------------------------
% ---------------------------------------------------------------------------------------------

\subsection{Vehicle Modelling}
\begin{figure}
    \centering
    \input{Figures/kinematic_model}
    \caption{Kinematic bicycle model for a tractor-trailer HVC.}
    \label{fig:vehdyn}
\end{figure}
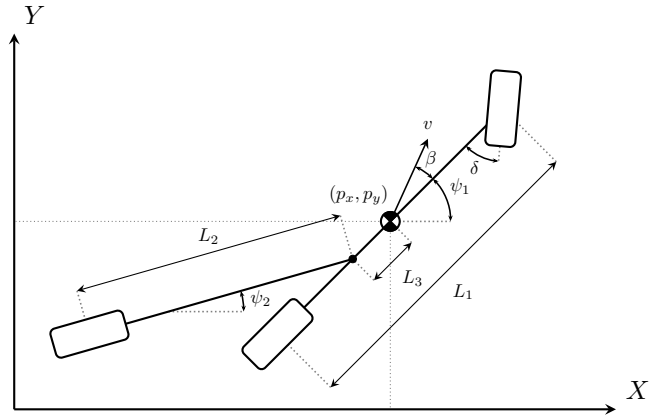
We model the dynamics of HVCs using an extended kinematic bicycle model, adapted from \cite{altafini2001some}, as follows.

\begin{equation*}%\label{eq:veh_dynamics}
\begin{split} 
            &\dot p_x = v\cos{(\psi_1 + \beta)}\\
            &\dot p_y = v\sin{(\psi_1 + \beta)}\\
            &\dot v = a \\
            &\dot \psi_1 = v \frac{\sin{{\beta}}}{L_1/2}\\
            &\dot \psi_2 = v \frac{\sin{(\psi_1 + \psi_2})}{L_2} - v \frac{(2L_3-L_1)\cos{(\psi_1 - \psi_2) \sin{(\beta)}}}{L_1 L_2}\\
            &\beta= \arctan(\tan (\delta)/2)            
\end{split}
\end{equation*}
The key features of the model are visualized in Fig. \ref{fig:vehdyn}. This includes: position of the center of the tractor unit in the inertial frame $p=(p_x,p_y)$; velocity $v$; tractor and trailer headings $\psi_1, \psi_2$; steering angle $\delta$, and lateral side-slip $\beta$. The shape of the vehicle is considered through the length of the tractor $L_1$, the distance from the end of the trailer to the connection point $L_2$ and the distance from the connection point to the center of the tractor $L_3$. The variables are concatenated to form the state $\xx = [p_x,\;p_y,\;v,\;\psi_1,\;\psi_2]^\top$, control input $\uu = [a,\; \delta]^\top$, and state dynamics $\dot \xx(\xx,\uu)$. For ease of notation, we consider the state and control input of all vehicles in the traffic scene as $\bxx = [\xx_\tx{e}^\top,\xx_1^\top, \dots, \xx_{N_V}^\top]^\top$ and $\buu = [\uu_{e}^\top,\uu_1^\top, \dots, \uu_{N_V}^\top]^\top$, respectively.

Indeed, the control actions of the ego-vehicle $\uu_\tx{e}$ will be described by the solution of an S-OCP, to be formulated later in Section~\ref{sec:S-OCP-formulation}. As in \cite{wang2023interaction, schuurmans2023safe}, control actions for human-driven HVCs will be described by a fixed set of control laws, each corresponding to an interactive decision. Importantly, the decision is considered as a discrete random variable $\xi \in \Xi$, where $\Xi = \{\tx{d}_1,\dots,\tx{d}_{N_\xi}\}$ describes a finite set of $N_{\xi}$ considered decisions. As human driver decisions may depend on the state of all vehicles in the traffic scene $\bxx$, we model the distribution of the random variable as follows.
\begin{equation} \label{eq:d_model}
    \xi \sim P_{\xi}(\bxx) = \left ( \mathbb{P}\{\xi = \tx{d} \;|\; \bxx\} \right )_{\tx{d} \in \Xi}
\end{equation}

We may then define a stochastic control law that yields the control action for each human driver.
\begin{equation}\label{eq:stoc_policy}
    \uu = \kappa(\bxx,\xi), \; \xi \sim P_{\xi}(\bxx)
\end{equation}

Finally, dynamics $\dot \bxx$ are combined with the control law \eqref{eq:stoc_policy} and discretized to produce the evolution of the traffic state,
\begin{equation}
    \bxx(k+1) = f(\bxx(k),\buu(k),\Bar{\xi}(k))
\end{equation}
where $\Bar{\xi}(k) = [\xi_1(k),\; \dots, \;\xi_{N_V}(k)]^\top$ describes the random decision variables for all human vehicles.

% ---------------------------------------------------------------------------------------------
% ---------------------------------------------------------------------------------------------

\subsection{Chance-Constrained Collision avoidance}
As we consider motion planning with small distance margins, we wish to adopt an accurate representation of the non-convex space that is occupied by the HVCs. To this end, we describe the space occupied by the ego-vehicle as a closed compact set $\mathbb{O}_{\tx{e}}(\xx_\tx{e})$ and, respectively, for the human-driven vehicles $\mathbb{O}_{\tx{i}}(\xx_i)$. We may then consider the squared Euclidean distance between the two sets as follows.
\begin{subequations} \label{eq:dist}
    \begin{align}
        \tx{dist}^2( \mathbb{O}_{\tx{e}}(\xx_\tx{e}), \mathbb{O}_{i}(\xx_{i}) ) \coloneqq \min_{p_\tx{e},p_{i}}\;  &||p_\tx{e} - p_{i}||_2^2 \\
        \tx{s.t.} \; &p_\tx{e} \in \mathbb{O}_{\tx{e}}(\xx_\tx{e}) \\
         &p_{i} \in \mathbb{O}_{i}(\xx_{i})
\end{align}
\end{subequations}
Indeed, the state $\xx_{i}$ and, furthermore, the set $\mathbb{O}_{i}(\xx_{i})$ may be uncertain as a consequence of uncertainty in human decisions. One may impose a constraint on \eqref{eq:dist} robustly, accounting for any possible state $\xx_i$ that has a non-zero probability $P_{\xi_i}(\bxx)$. However, this can yield an excessively conservative control strategy when some decisions are highly unlikely. Instead, we adopt chance constraints to allow constraint violations in these cases. We may express a general chance-constraint on collision avoidance between the two vehicles as,
\begin{equation} \label{eq:cc_collisions}
    \mathbb{P}_{\zz} \Big \{ \tx{dist}^2(\mathbb{O}_{\tx{e}}(\xx_\tx{e}),\mathbb{O}_{i}(\xx_i) \geq d^2_\tx{safe} \Big \} \geq 1 - \varepsilon
\end{equation}
where $d_\tx{safe}$ is a positive safety margin and $\varepsilon \in [0,1]$ represents the maximum allowed probability of violating the constraint. Here $\mathbb{P}_\zz$ is a probability measure with respect to a random variable $\zz$. For now, we will keep this definition general and later expand on the important implications of the choice of probability measure in Section \ref{sec:CCs_in_trees}.

% ---------------------------------------------------------------------------------------------
% ---------------------------------------------------------------------------------------------

\subsection{Stochastic Optimal Control Problem} \label{sec:S-OCP-formulation}

We now introduce the last standard ingredients and define the discrete-time S-OCP. To treat physical limitations of the ego-vehicle, we introduce bounds on the state and control variables, here noted as vector-valued functions ${h(\xx_\tx{e},\uu_\tx{e}) \leq 0}$, and $h_N(\xx_\tx{e}) \leq 0$. 
We consider a quadratic stage cost as,
\begin{equation} \label{eq:stage_cost}
    \ell(\xx_\tx{e},\uu_\tx{e}) = ||\xx_\tx{e} - \xx_{\mathrm{ref}}||_\mathbf{Q}^2 +||\uu_\tx{e}||_\mathbf{R}^2 +||\Delta \uu_\tx{e}||_{\mathbf{R}_\Delta}^2
\end{equation}
where $\mathbf{Q} \succeq 0, \;\mathbf{R} \succ 0,\mathbf{R}_\Delta \succ 0$ are cost matrices and $\xx_{\mathrm{ref}}$ is a reference state trajectory. Change in control actions $\Delta \uu_\tx{e}(k) = \uu_\tx{e}(k)-\uu_\tx{e}(k-1)$ is included to consider comfort, without the need for additional state variables.  We further introduce a terminal cost $\ell_N(\xx_\tx{e}(N)) = ||\xx_\tx{e} - \xx_{\mathrm{ref}}||_\mathbf{P}^2$, where $\mathbf{P} \succeq 0$.

Finally, the complete chance-constrained S-OCP becomes,
\begin{subequations}\label{eq:socp}
    \begin{align} \label{eq:socp_cost}
        \min_{\xx_\tx{e},\uu_\tx{e}} & \mathbb{E}_{\Bar{\xi}} \left [\;\ell_N(\xx_\tx{e}(N)) + \sum_{k=0}^{N-1} \ell \left (\xx_\tx{e}(k), \uu_\tx{e}(k) \right ) \right ]\\ \label{eq:socp_eq}
        \tx{s.t.},& \; \bxx(k+1) = f(\bxx(k),\buu(k),\Bar{\xi}(k)) \\\label{eq:socp_ineq}
        &h(\xx_\tx{e}(k),\uu_\tx{e}(k)) \leq 0, \;h_N(\xx_\tx{e}(N)) \leq 0\\ \label{eq:socp_policy}
        & \uu_{i}(k) = \pi_{i}(\bxx(k),\xi_{i}(k)), \; \xi_{i}(k) \sim P_{\xi_{i}}(\bxx(k))\\ \label{eq:socp_collisions}
        &\mathbb{P}_{\zz} \Big \{ \tx{dist}^2(\mathbb{O}_{\tx{e}}(\xx_\tx{e}),\mathbb{O}_{i}(\xx_i)) \geq d^2_\tx{safe} \Big \} \geq 1 - \varepsilon\\
        &\bxx(0) = \bxx(t)
    \end{align}
\end{subequations}
where \eqref{eq:socp_eq}, \eqref{eq:socp_policy}, $h(\xx_\tx{e}(k),\uu_\tx{e}(k))$ are imposed for $k = 0,\dots, N-1$, and \eqref{eq:socp_collisions} is imposed for $k = 0,\dots, N$. Further, \eqref{eq:socp_policy} and \eqref{eq:socp_collisions} are repeated for all uncontrollable vehicles $i$.
The initial condition $\bxx_\tx{e}(t)$ is the continuous traffic state, measured at the current time $t$.

Indeed, this problem is intractable for gradient-based numerical optimization solvers as: (i) The decisions of the surrounding vehicles $\Bar{\xi}$ are stochastic, with an unknown probability distribution; (ii) The squared distance function \eqref{eq:dist} is non-smooth and includes non-convex obstacles; (iii) The chance constraint on the squared distance \eqref{eq:cc_collisions} is non-smooth. In Section \ref{sec:scenario_trees} we address (i) by optimizing over a scenario tree, and in Section \ref{sec:constraint_reform} we address (ii, iii) by proposing tight outer-approximations.

% ---------------------------------------------------------------------------------------------
% ---------------------------------------------------------------------------------------------
% ---------------------------------------------------------------------------------------------
% ---------------------------------------------------------------------------------------------
% ---------------------------------------------------------------------------------------------

\section{Scenario Tree Reformulation} \label{sec:scenario_trees}
As there are a finite number of possible decisions that human drivers can make at each discrete time step $k$, it is possible to enumerate the combinations of all decisions and, further, the possible future states at time $k+1$. Repeating this procedure over the prediction horizon, from $k=0$ to $k=N$, yields a tree structure of all possible states and control inputs of human drivers over the horizon. A direct consequence is that the multimodal behavior observed in humans is encoded in the structure of the decision tree. Alternative prediction methods often consider the most probable path through the tree, at the expense of losing this multimodality. In the following subsections, we formally introduce the construction of the scenario tree for our S-OCP, together with a method for learning the decision distribution $\xi \sim P_\xi(\bxx)$. The theory presented here is applicable for any finite number of human vehicles and decisions, but we limit the analysis here, and later in the simulation study, to one human vehicle. Hence, we opt to utilize the subscript $\tx{h}$ to refer to the human-driven vehicle.

% ---------------------------------------------------------------------------------------------
% ---------------------------------------------------------------------------------------------

\subsection{Scenario Tree Construction}
The scenario tree is constructed from its root, consisting of the current traffic state at time $t$, and is expanded until it terminates at its leaf nodes. To describe the evolution of the scenario tree, we adapt the notation of \cite{wang2020non,chen2022interactive} to our setting. Each node is indexed by $\iota$, where $\iota \in \mathbb{N}=\{0,\dots,N_\iota\}$ is the set of all nodes in the scenario tree. For notational convenience, we also define $\iota \in \NN_k$ as all nodes at time step $k$. We use $\tx{Ch}(\iota)$ to denote the set of children and $\tx{Pre}(\iota)$ to denote the parent of a node $\iota$. To describe transitions, we introduce $\iota^+$ as the directly consecutive nodes from a node $\iota$, i.e., $\iota^+ \in \tx{Ch}(\iota)$, $\iota=\tx{Pre}(\iota^+)$. To restrict the size of the tree, we adopt the method in \cite{chen2022interactive} and limit the number of nodes where all possible decisions of the human vehicle are considered. This yields a fixed, relatively small set of ``\textit{branching nodes}'' $\iota \in \mathbb{N}_{\tx{br}}$, where all possible human decisions are enumerated to expand the tree. We similarly define ``\textit{non-branching}'' nodes $\iota \notin \mathbb{N}_{\tx{br}}$ as nodes where the decision of the human is inherited by that of its parent. By convention, we always consider the root node as a branching node, while the number and location of other branching nodes become parameters that are fixed apriori.

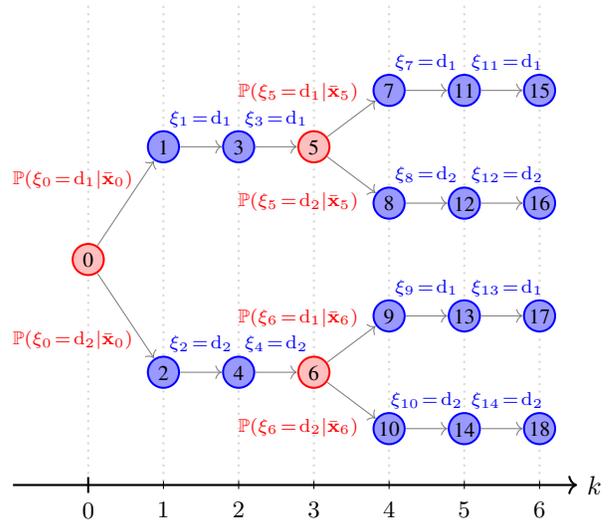
\begin{figure}
    \centering
        \input{Figures/scenario_tree}
    \caption{An example of a scenario tree approximation over a horizon $N=6$ based on ${N_\xi = 2}$ decisions $\tx{d}$ of an uncontrollable vehicle. Branching and non-branching nodes are marked with red and blue, respectively. The different realizations of the random variable $\xi_\iota$ are similarly indicated in the corresponding color at each node.}
    \label{fig:scenario_tree}
\end{figure}

An example of such a scenario tree, generated by a single uncontrollable human vehicle with $\xi \in \Xi = \{\tx{d}_1,\tx{d}_2\}$, is illustrated in Fig. \ref{fig:scenario_tree}. Here we, e.g., have $\NN = \{0,\dots, 18\}$, $\NN_\tx{br} = \{0,5,6\}$ and $\NN_{N} = \{15,16,17,18\}$ for the leaf nodes at $k=N$. Note that the non-branching nodes adopt the same decision as that of their parent and only have one child node. 

\begin{remark} Indeed, these scenario trees are approximations of the true underlying Markov process, where one may consider each node as a branching node. However, as the number of nodes and, furthermore, the number of optimization variables scales exponentially with the length of the horizon $N$, problems quickly become intractable. Practical applications typically employ scenario reduction techniques; see, e.g., \cite{jacobsen2025combined}.
\end{remark}

% % ---------------------------------------------------------------------------------------------
% % ---------------------------------------------------------------------------------------------

\subsection{Optimal Control over Scenario Trees}
Each node $\iota$ in the scenario tree contains the complete traffic state $\bxx_\iota$, and each node, excluding leaf nodes, also contains the ego-vehicle control actions $\uu_\tx{e,\iota}$ and random variables $\xi_\iota$. The realizations of the random variables describe transitions $\iota\to \iota^+$, for every  ${\iota \in \NN \! \setminus \! \NN_N}$. With slight abuse of notation, we refer to $\tx{d}_{\iota^+}$ as the human decision at node $\iota$ that results in a transition to node $\iota^+$. Hence, we may express the state dynamics over the scenario tree as,
\begin{equation}
    \bxx_{\iota^+} = f(\bxx_\iota,\buu_\iota|\xi_\iota = \tx{d}_{\iota^+}), \; \forall \iota^+ \in \tx{Ch}(\iota).
\end{equation}
The dynamics propagate through the tree until the leaf nodes are reached, i.e., they are applied at all nodes ${\iota \in \NN \! \setminus \! \NN_N}$. 
At each node $\iota$, we can express the transition probability to a node $\iota^+$ based on the decision distribution $\xi_\iota \sim P_{\xi_\iota}(\bxx_\iota)$. To treat non-branching nodes i.e., nodes with a single child, we express this transition probability as follows.

\begin{equation} \label{eq:P_transition}
    P_{\iota \to \iota^+}(\bxx,\tx{d}_{\iota^+}) = \begin{cases}
        P_{\tx{d}_{\iota^+}}(\bxx_\iota), & \iota \in \NN_\tx{br} \\
        1.0, & \iota \notin \NN_\tx{br}
    \end{cases}
\end{equation}
Here, the case $\iota \notin \NN_\tx{br}$ describes transitions from non-branching nodes. Further, we can apply \eqref{eq:P_transition} recursively to express the probability of transitioning from some node $j$, to another reachable node $\iota^+$ as,
\begin{equation} \label{eq:P_transition_recursive}
    p_{j \to \iota^+} = p_{j \to \iota} P_{\iota \to \iota^+}(\bxx_\iota,\tx{d}_{\iota^+})
\end{equation}
where we consider the trivial initial condition as $p_{j \to j} =1.0$. We can now reformulate the expected cost \eqref{eq:socp_cost} with respect to the transition probabilities, as a function of the ego-vehicle state and control ($\xx_\tx{e,\iota}$, $\uu_\tx{e,\iota}$) over the scenario tree,
\begin{equation}
    \begin{split}
    J(\xx_\tx{e}, \uu_\tx{e})  = & \sum_{k=0}^{N-1}\sum_{\iota \in \NN_k}  p_{0 \to \iota} \ell(\xx_{\tx{e},\iota},\uu_{\tx{e},\iota})\\
    & + \sum_{\iota \in \NN_N}  p_{0 \to \iota} \ell_N(\xx_{\tx{e},\iota}).
    \end{split}
\end{equation}
Note that the leaf nodes $\iota \in \NN_N$ are treated separately from the others, to properly assign the terminal cost. To treat physical limitations of the ego-vehicle, we introduce bounds on the deterministic ego-state and control variables, again noted as vector-valued functions ${h(\xx_{\tx{e},\iota}\uu_{\tx{e},\iota}) \leq 0},\; \forall \iota \in \NN \! \setminus \! \NN_N$, and $h_N(\xx_{\tx{e},\iota}) \leq 0, \; \forall \iota \in  \NN_N$.

% % ---------------------------------------------------------------------------------------------
% % ---------------------------------------------------------------------------------------------

\subsection{Chance Constraints in Scenario Trees} \label{sec:CCs_in_trees}
Lastly, we wish to formulate the chance-constrained collision avoidance \eqref{eq:cc_collisions}, with respect to different probability measures.
As a consequence of the definition of the random variable $\xi$, the uncertainty in the formulated problem is entirely confined to the transitions between the nodes in the tree. Indeed, it is possible to construct multiple valid chance constraints depending on which transitions and, consequently, which combination of random variables you wish to consider. Naturally, the guarantees provided by each chance constraint only apply to the transitions considered. Here, we present three valid chance constraints alternatives: ``\textit{Joint}'', considering every transition through the tree jointly; ``\textit{Stage-based}'', considering transitions to each time step $k$ separately, and ``\textit{Node-based}'' adapted from \cite{wang2023interaction}, considering transitions from each node separately.

\subsubsection{Joint Chance Constraints}
In the joint setting, we wish to express and constrain the probability of encountering any constraint violations in the tree. We may express this as follows.
\begin{equation*}
    \mathbb{P}_{\xi_0,\dots,\xi_{N_\iota}} \! \Big \{ \! \bigcup_{\iota \in \mathbb{N}} \tx{dist}^2(\mathbb{O}_{\tx{e}}(\xx_\tx{e,\iota}),\mathbb{O}_{i}(\xx_{\tx{h},\iota}) \leq d^2_\tx{safe} \big | \bxx_0 \Big \} \! < \varepsilon
\end{equation*}
The following proposition provides an outer approximation of the above constraint.
\begin{proposition} \label{prop:joint_cc}
    Consider a scenario tree defined over a discrete-time prediction horizon $k=0,\dots,N$ with nodes $\iota \in \mathbb{N}=\{\mathbb{N}_0,\dots,\mathbf{N_k},\dots,\mathbb{N}_N\}$, containing states $\xx_\iota$. Further, consider random variables $\zz_\iota \in \mathbb{Z}=\{0,1, \dots,{N_\zz}\}$ whose distribution $\zz_\iota \sim P(\xx_\iota) =\left ( \mathbb{P}\{\zz_\iota = i \;|\; \xx_\iota\} \right )_{i \in \mathbb{Z}}$ defines the transition probabilities $p_{\iota \to \iota^+}$, where $\iota^+ \in \tx{Ch}(\iota)$. 
    For a state constraint $g(\xx_{\iota^+}) \leq 0$, where $g:\RR^{n_x}\mapsto\RR$, we may then constrain the joint probability of constraint violations as follows.
    \begin{align} \label{eq:CC_joint_exact}
    &  \mathbb{P}_{\zz_0,\dots \zz_{N_\iota}} \Big \{ \bigcup_{\iota \in \mathbb{N}} g(\xx_\iota) > 0 \, \big | \, \bxx_0 \Big \} \leq \\\label{eq:CC_joint}
    & \sum_{k=0}^{N-1} \sum_{ \iota^+ \in \NN_{k+1}} p_{0\to\iota^+} \mathbf{1}_{(0,\infty)} \big (g(\xx_{\iota^+})\big) < \varepsilon
    \end{align}
Here, $\varepsilon \in [0,1]$ and $\mathbf{1}_{(0,\infty)}$ is an indicator function of the positive real line.
\end{proposition}
\textit{Proof:} Straightforward adaptation of \cite{ono2015chance} for the scenario tree setting; see Appendix \ref{A:proof_joint_cc}.

\begin{remark}
    The constraint \eqref{eq:CC_joint} indeed sums over all nodes in the scenario tree and is added once in the optimization problem. It provides an outer approximation of \eqref{eq:CC_joint_exact} as, instead of constraining the probability of the existence of constraint violations, it constrains \textit{the total expected number of constraint violations}. We will later compute this metric in the simulation study.
\end{remark}

\subsubsection{Stage-based Chance Constraints}
In this setting, the chance constraints treat transitions between each discrete time step separately. With nodes at discrete time $k$, $\iota \in \mathbb{N}_k$, and children $\iota^+\in\tx{Ch}(\iota)$ we may express and reformulate the constraint as follows.
\begin{align} \nonumber
    & \mathbb{P}_{\{ \zz_0,\dots,\zz_\iota \, \forall \iota \in \mathbb{N}_k \}} \Big [\bigcup_{\iota^+ \in \mathbb{N}_{k+1}} g(\xx_{\iota^+}) > 0 \, \big | \, \xx_0 \Big ]< \varepsilon \Leftrightarrow \\ \label{eq:CC_stage_based}
    &\sum_{\iota^+ \in \mathbb{N}_{k+1}} p_{0\to \iota^+} \mathbf{1}_{(0,\infty)} \left (g(\xx_{\iota^+}) \right) < \varepsilon
\end{align}
The proof follows immediately from the proof of Proposition \ref{prop:joint_cc}. The constraints are repeated for each $k=0,\dots,N-1$ and constrain the expected number of collisions for the transitions between each time step, respectively.

\subsubsection{Node-based Chance Constraints}
Finally, we wish to formulate chance constraints on the transitions from each node, i.e., transitions $i\to \iota^+, \forall \iota^+ \in \tx{Ch}(\iota)$. Crucially, we wish to consider an approximated scenario tree where only a subset of nodes $j \in \mathbf{N_\tx{br}} \subset \mathbb{N}$ expand the scenario tree. Hence, nodes $\iota^+ \in \tx{Ch}(\tx{Ch}(\dots\tx{Ch}(j)))$ need to consider the transition probability from a branching parent $j$ directly. To simplify notation, we introduce $\tx{Pre}_\tx{br}(\iota^+)$, as the most recent branching parent of a node $\iota^+$. We may express the chance constraints over such transitions as follows.
\begin{align} \nonumber
        &\mathbb{P}_{\zz_j} \Big [\bigcup_{\iota^+ \in \tx{Ch}(\tx{Ch}(\dots\tx{Ch}(j)))} g(\xx_{\iota^+}) > 0 \, \big | \,\xx_j \Big ] < \varepsilon, \Leftrightarrow \\ \label{eq:CC_node_based}
        &\sum_{\iota^+ \in \tx{Ch}(\tx{Ch}(\dots\tx{Ch}(j)))} p_{j\to\iota^+} \mathbf{1}_{(0,\infty)} \left (g(\xx_{\iota^+}) \right)< \varepsilon
\end{align}
Here,$j=\tx{Pre}_\tx{br}(\iota^+)$. Similarly, the proof follows immediately from the proof of Proposition \ref{prop:joint_cc}. The constraint \eqref{eq:CC_node_based}, is repeated for all branching nodes $j \in \mathbb{N}_\tx{br}$ and for all of it's children $\iota^+ \in \tx{Ch}(\tx{Ch}(\dots\tx{Ch}(j)))$ such that $j=\tx{Pre}_\tx{br}(\iota^+)$.

% % ---------------------------------------------------------------------------------------------
% % ---------------------------------------------------------------------------------------------

\subsection{Estimating Decision Distribution}
In our setting, the decision distribution $P_\xi(\bxx_\iota)$ is unknown and requires estimation. Importantly, the estimation must be dependent on the traffic state $\bxx$, so that interactions between vehicles can be directly accounted for in the optimal control problem. Following \cite{wang2023interaction}, we estimate the conditional distribution given a state $\bxx_\iota$ and a decision $\tx{d}$ using logistic regression as
\begin{equation} \label{eq:p_hat}
    \hat{P}(\bxx_\iota, \tx{d};{\boldsymbol{\theta}}) = \left( \frac{\exp{\left( \theta_\tx{d}^\top \varphi(\bxx_\iota) \right)}}{\sum_{\tx{d} \in \Xi}\exp{\left( \theta_\tx{d}^\top \varphi(\bxx_\iota) \right)}} \right)
\end{equation}
where $\boldsymbol{\theta}= [\theta_{\tx{d}_1},\dots,\theta_{\tx{d}_{N_{\xi}}}] \in \mathbb{R}^{N_\theta \times N_\xi}$ are learnable weights and $\varphi(\bxx) \in \mathbb{R}^{N_\theta}$ is a feature vector, constructed based on the traffic state $\bxx$.  In the final S-OCP, \eqref{eq:p_hat} is utilized in \eqref{eq:P_transition} to estimate the transition probability from branching nodes. Hence, we obtain an estimate of the transition probability as $\hat{P}_{\iota \to \iota^+}(\bxx_\iota,\tx{d}_{\iota^+};\boldsymbol{\theta})$.

\section{Tractable S-OCP Reformulation} \label{sec:constraint_reform}
 The squared distance formulation with non-convex sets \eqref{eq:dist} and the chance constraints, \eqref{eq:CC_joint}, \eqref{eq:CC_stage_based}, \eqref{eq:CC_node_based} are not continuously differentiable. Further, it is important to consider tight reformulations of these constraints, as the distance margins are small. In this section, we present computationally efficient options for such reformulations and present the complete tractable S-OCP.

\subsection{Squared Distance Reformulation}
Indeed, the union of an infinite number of closed compact convex sets can represent any closed compact set \cite{ziegler2012lectures}. However, tractable OCP formulations are restricted to a small number of convex sets, as the number of collision avoidance constraints scales exponentially with the number of sets used to represent each vehicle. As indicated by previous work, convex polytopes are an attractive choice for the convex set, as they provide an accurate vehicle representation, with a small number of sets \cite{zhang2020optimization,dietz2023efficient,fan2023efficient}. Hence, we opt to utilize the union of convex polytopes as an accurate representation of the HVCs. Let $\OO(\xx) = \{\bigcup_l \OO_{l}(\xx) \}$ be the union of $N_{l}$ convex polytopes $\OO_{l}(\xx) \in \mathbb{R}^n$ representing the occupied space of a vehicle in $n$, where the index $l\in \mathbf{I}_{N_{l}} = \{1,\dots, N_{l} \}$. Ensuring collision avoidance between two vehicles $\tx{e}$ and $i$ amounts to ensuring that none of their respective sets $l$ and $\tilde{l}$ intersect. Hence, we extend \eqref{eq:dist} to consider multiple pairs of sets, with a single scalar constraint,
\begin{equation} \label{eq:dist_shortest}
    \min_{(l,\tilde{l}) \in \, \mathbf{I}_{N_{l,\tx{e}}} \times \,\mathbf{I}_{N_{\tilde{l},i}}} \tx{dist}^2(\mathbb{O}_{\tx{e},l}(\xx_\tx{e}),\mathbb{O}_{i,\tilde{l}}(\xx_i))  \geq d^2_\tx{safe} 
\end{equation}
where the Cartesian product $\mathbf{I}_{N_{l,\tx{e}}} \! \!\times \! \mathbf{I}_{N_{\tilde{l},i}}$ represents the set of all $N_c$ possible combinations of indices $l$ and $\tilde{l}$. I.e., the above problem returns the smallest squared distance between the sets $\OO_\tx{e}(\xx_\tx{e})$ and $\OO_i(\xx_i)$ by finding the closest sets $l,\tilde{l}$ and enforces a strictly positive safety margin $d_\tx{safe}> 0$. In this work, we will consider $N_{l}=2$ sets in dimension $n=2$, and one human-driven vehicle $\tx{h}$ but it is straightforward to extend our collision avoidance approach to any number of sets, vehicles, and spatial dimensions. 

To describe each polytopic set $l$, we consider $\mathcal{V}$-polytopes, defined by a convex combination of vertices $V_{l} \in \mathbb{R}^{n \times N_v}$, where $N_v$ represents the number of vertices of a set with index $l$ that partially describes a vehicle. Choosing $\Tilde{V}_{l}$ to describe the vertices when the point mass of the vehicle is located at the origin, we can represent any configuration of the convex set in $\mathbb{R}^n$ using a rotation matrix $R(\xx)\in \mathbb{R}^{n \times n}$ and a translation vector $t(\xx)\in \mathbb{R}^{n \times N_v}$ as $V_{l}(\xx) = R(\xx) \Tilde{V}_{l} + t(\xx)$. The occupied space of a set $l$, which partially describes the vehicle, then becomes,
\begin{equation}\label{eq:V_rep}
    \mathbb{O}_{l}(\xx) = \{V_{l}(\xx)\phi_{l} \; : \; \mathbf{1}^\top \phi_{l} = 1, \phi_{l} \geq \mathbf{0}\}
\end{equation}
where $\phi_{l} \in \mathbb{R}^{N_v}$ and $\mathbf{1}$ is a column vector with ones in all entries \cite{ziegler2012lectures}. Utilizing duality theory from convex optimization as in \cite{dietz2023efficient}, we may reformulate the squared distance between two sets $l$ and $\tilde{l}$ that partially describe the ego-vehicle $\tx{e}$ and a human-driven vehicle $\tx{h}$, respectively.

\begin{proposition} \label{prop:dist2_dietz}
    Let $\xx_\tx{e}$ and $\xx_\tx{h}$ represent the states of vehicles $\tx{e}$ and $\tx{h}$ and consider sets $\OO_{\tx{e},l}(\bxx_\tx{e})$ and $\OO_{\tx{h},\tilde{l}}(\bxx_\tx{h})$ described using \eqref{eq:V_rep}. We then have the following equivalence,
    \begin{align} \nonumber
        & \tx{dist}^2\big(\OO_{\tx{e},l}(\xx_\tx{e}),\OO_{\tx{h},\tilde{l}} (\xx_\tx{h})\big) \geq d^2_\tx{safe} \Longleftrightarrow \\ \label{eq:thm_dist}
        & \exists \zeta \in \mathbb{R}^n, \mu,\nu \in \mathbb{R}: -\frac{1}{4}\zeta^\top\zeta - \mu - \nu \geq d^2_\tx{safe}\\ \label{eq:thm_multipliers}
        &V_{\tx{e},l}(\xx_\tx{e})^\top\zeta + \mu \mathbf{1} \geq \mathbf{0},\; -V_{\tx{h},\tilde{l}}(\xx_\tx{h})^\top\zeta + \nu \mathbf{1} \geq \mathbf{0} 
    \end{align}
    for some $d_\tx{safe}>0$.
\end{proposition}
\textit{Proof:} By additionally describing the obstacles using \eqref{eq:V_rep}, 
 \eqref{eq:dist} becomes a convex optimization problem in $\phi_{\tx{e},l},\phi_{\tx{h},\tilde{l}}$. Hence, by also representing the obstacles using \eqref{eq:V_rep}, the proof follows immediately from \cite[Proposition~1]{dietz2023efficient}.

We may now consider the case where the shape of each vehicle is represented with multiple convex polytopes. In this setting, we extend Proposition \ref{prop:dist2_dietz} to express a constraint on the shortest distance between two vehicles, each represented with multiple polytopes.
\begin{corollary} \label{cor:shortest_distance}
    Consider the squared-distance reformulation of Proposition \ref{prop:dist2_dietz} for multiple sets $\OO_{\tx{e},l}(\bxx_\tx{e})$ and $\OO_{\tx{h},\tilde{l}}(\bxx_\tx{h})$ defined as \eqref{eq:V_rep}, where $l\in\mathbf{I}_{N_{l,\tx{e}}} = \{1,\dots, N_{l,\tx{e}} \}$, $\tilde{l}\in \mathbf{I}_{N_{\tilde{l},\tx{h}}} = \{1,\dots, N_{\tilde{l},\tx{h}} \}$. We then have the following equivalence,
    \begin{align} \nonumber
        &\min_{(l,\tilde{l}) \in \, \mathbf{I}_{N_{l,\tx{e}}} \times \,\mathbf{I}_{N_{\tilde{l},\tx{h}}}} \tx{dist}^2(\mathbb{O}_{\tx{e},l}(\xx_\tx{e}),\mathbb{O}_{\tx{h},\tilde{l}}(\xx_\tx{h}))  \geq d^2_\tx{safe} \Leftrightarrow\\ \label{eq:cor1_shortest_dist}
        &\exists \gamma \in \RR, \zeta \in\RR^{n \times N_c},\mu, \nu \in \RR^{N_c}: \gamma \leq -d^2_\tx{safe} \\ \label{eq:cor1_multipliers}
        & \begin{cases}
             \frac{1}{4}\zeta_{(l,\tilde{l})}^\top\zeta_{(l,\tilde{l})} + \mu_{(l,\tilde{l})} + \nu_{(l,\tilde{l})} \leq \gamma \\
             V_{\tx{e},l}(\xx_\tx{e})^\top\zeta_{(l,\tilde{l})} + \mu_{(l,\tilde{l})} \mathbf{1} \geq \mathbf{0}\\
             -V_{\tx{h},\tilde{l}}(\xx_\tx{h})^\top\zeta_{(l,\tilde{l})} + \nu_{(l,\tilde{l})} \mathbf{1} \geq \mathbf{0}
        \end{cases}\nquad
    \end{align}

where $d_\tx{safe}\!>\!0$ and \eqref{eq:cor1_multipliers} are repeated $\forall (l,\hat{l}) \!\!\in\! \mathbf{I}_{N_{l,\tx{e}}} \! \!\times \mathbf{I}_{N_{\hat{l},\tx{h}}}$.
\end{corollary}
\textit{Proof}: See Appendix \ref{A:proof_shortest_d}.

To ease notation, we rearrange the constraints \eqref{eq:cor1_multipliers}, without loss of generality, and define them jointly as
\begin{equation}
    \boldsymbol{\Gamma}(\bxx,\gamma, \zeta,\mu, \nu) \leq 0.
\end{equation}
 The dual objective function \eqref{eq:cor1_shortest_dist} now expresses the constraint on the squared distance between the possibly non-convex sets $\mathbb{O}_{\tx{e}}(\xx_\tx{e})$, $\mathbb{O}_{\tx{h}}(\xx_\tx{h})$. For future reference, we refer to the additional optimization variables $\zeta,\mu$ and $\nu$ as ``\textit{collision multipliers}'' and treat $\gamma$ as a tight approximation of the smallest squared distance, with a negative sign.

 \subsection{Tight Chance Constraint Reformulation}

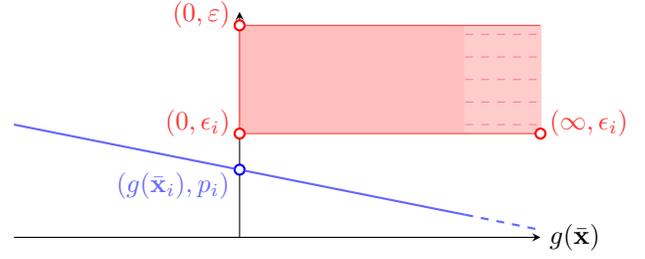
\begin{figure}
    \centering
    \input{Figures/cc_reform}
    \caption{Conceptual visualization of the tight chance constraint approximation for one combination of $\epsilon_i$, $p_i$ and $\bxx_i$. The infeasible set is displayed in red and the constraint \eqref{eq:prop1_ccs} is displayed in blue for some $\lambda_i$.}
    \label{fig:cc_reform}
\end{figure}

Following Corollary \ref{cor:shortest_distance} we may express the shortest distance constraint between two vehicles $\tx{e}$ and $\tx{h}$ as a scalar-valued  constraint ${g(\bxx) = \gamma +d^2_\tx{safe} \leq 0}$, where the state dependence stems from the connection of $\gamma$ with the constraints \eqref{eq:cor1_multipliers}. Our three chance constraint variants \eqref{eq:CC_joint}, \eqref{eq:CC_stage_based}, \eqref{eq:CC_node_based} all consist of scaled summations of indicator functions. Indeed, this formulation is poorly suited for numerical optimization as the indicator functions are not continuously differentiable. Common outer approximations in this setting include, e.g., the Conditional Value at Risk \cite{nemirovski2012safe} or sigmoid functions \cite{wang2023interaction}. However, neither approach offers tight guarantees, posing challenges for motion planning with small collision margins, i.e., when $g(\bxx)$ approaches zero. To this end, we propose a tight outer-approximation with the following proposition.

\begin{proposition} \label{prop:cc_reform}
    Consider a vector $\pp \in \RR^m$ such that $\sum_i p_i = 1$, $p_i \in [0,1]$ with corresponding constraints $g(\bxx_i)\leq 0$ where $g:\RR^{n_x} \mapsto \RR$ and $\bxx_i \in \RR^{n_x}$. We then have the following equivalence.

    \begin{align}\nonumber
        & \sum_{i=1}^{m} p_i \mathbf{1}_{(0,\infty)} \big ( g(\bxx_{i}) \big)  \leq \varepsilon \Longleftrightarrow\\ \label{eq:prop1_sum}
        &\exists \epsilon \in \RR^{m},\lambda \in \RR^{2\times m}: \sum_{i=1}^m \epsilon_{i} \leq \varepsilon\\ \label{eq:prop1_ccs}
        &\begin{cases}
        \lambda_{1,i}g(\bxx_{i}) + \lambda_{2,{i}}\!\left(p_i - \epsilon_{i}\right) \! < 0 \\
        \lambda_{1,{i}} > 0,\;\lambda_{2,{i}} > 0\;, \epsilon_{i} \geq 0
        \end{cases} \ncol\ncol\ncol
    \end{align}
where $\varepsilon \in [0,1]$ and \eqref{eq:prop1_ccs} is repeated $\forall {i}\in \{1,\dots,m \}$.
\end{proposition}
\textit{Proof:} See Appendix \ref{A:proof_cc_reform}.

A conceptual visualization of \eqref{eq:prop1_ccs} is displayed in Fig. \ref{fig:cc_reform}. Indeed, this form is directly applicable to the three different chance constraint versions \eqref{eq:CC_joint}, \eqref{eq:CC_stage_based}, \eqref{eq:CC_node_based}, as it holds for any such summation and $\varepsilon_i \in [0,\varepsilon]$, $p_i \in [0,1]$, as in our setting. Importantly, each version $\tx{v} \in \{\text{joint},\text{ stage-based},\text{ node-based}\}$ considers variables defined for different transitions between different sets of nodes $\mathbb{N}_\tx{v}$, as defined in \eqref{eq:CC_joint}, \eqref{eq:CC_stage_based}, \eqref{eq:CC_node_based}. For brevity, we rearrange the constraints \eqref{eq:prop1_sum}, \eqref{eq:prop1_ccs}, for each $\tx{v}$ as,
\begin{equation}
    \Psi_\tx{v}(\mathbb{N}_\tx{v};\varepsilon) \leq 0
\end{equation}
and refer to \eqref{eq:CC_joint}, \eqref{eq:CC_stage_based}, \eqref{eq:CC_node_based} for the rigorous definitions. For future reference, we refer to the additional optimization variables $\epsilon$ and $\lambda$ as ``\textit{chance constraint multipliers}''.

\subsection{Tractable S-OCP Formulation}
Combining the objective, dynamics, and distribution estimation in Section \ref{sec:scenario_trees} with the reformulated distance and chance constraints in Section \ref{sec:constraint_reform} we may now express the complete, reformulated S-OCP,
\begin{subequations} \label{eq:S-OCP_reform}
    \begin{align} \label{eq:OCP_objective}
        &\ncol\ncol\ncol\min_{\substack{\xx_\tx{e},\uu_\tx{e}\\\boldsymbol{\gamma},\boldsymbol{\lambda}, \boldsymbol{\epsilon}, \boldsymbol{\zeta},\boldsymbol{´\mu},\boldsymbol{\nu}}} \;&& \ncol \ncol \ncol \ncol \ncol \ncol J(\xx_\tx{e},\uu_\tx{e})\\ \label{eq:OCP_dynamics}
        & \;\; \tx{s.t.}\; && \ncol \ncol \ncol \ncol \ncol \ncol \bxx_{\iota^+} = f(\bxx_\iota,\buu_\iota|\xi_\iota = \tx{d}_{\iota^+})\\ \label{eq:OCP_p_transition}
        & && \ncol \ncol \ncol \ncol \ncol \ncol p_{0 \to \iota^+} = p_{0 \to \iota} \hat{P}_{\iota \to \iota^+}(\bxx,\tx{d}_{\iota^+};\boldsymbol{\theta})\\ \label{eq:OCP_boxconstraints}
        & && \ncol \ncol \ncol \ncol \ncol \ncol h(\xx_\iota, \uu_\iota) \leq 0, \;h_N(\xx_\iota) \leq 0\\ \label{eq:OCP_dist_reformed}
        & && \ncol \ncol \ncol \ncol \ncol \ncol \boldsymbol{\Gamma}(\bxx_\iota,\gamma_\iota, \zeta_\iota,\mu_\iota, \nu_\iota) \leq 0\\ \label{eq:OCP_cc_reformed} 
        & && \ncol \ncol \ncol \ncol \ncol \ncol \Psi_\tx{v}(\mathbb{N}_\tx{v};\varepsilon) \leq 0\\ \label{eq:OCP_initial_cond}
        & && \ncol \ncol \ncol \ncol \ncol \ncol  \bxx_0 = \bxx(t).
    \end{align}
\end{subequations}
The evolution of the state and transition probability \eqref{eq:OCP_dynamics}, \eqref{eq:OCP_p_transition} are imposed $\forall \iota^+ \in \tx{Ch}(\iota)$ where ${\iota \in \NN \! \setminus \! \NN_N}$. The ego-vehicle constraints $h$ are enforced $\forall \iota \in \mathbb{N} \setminus \mathbb{N}_N$ with terminal constraints $h_N$, enforced $\forall \iota \in \mathbb{N}_N$. The constraints on the collision multipliers \eqref{eq:OCP_dist_reformed} are enforced $\forall \iota \in \mathbb{N}$. The tight outer approximation of the chance constraints on the squared distance \eqref{eq:OCP_cc_reformed}, is enforced $\forall \iota \in \mathbb{N}$ and importantly varies with version $\tx{v} \in \{\text{joint},\text{ stage-based},\text{ node-based}\}$. Indeed, each formulation considers a different set of transitions between a different set of nodes $\mathbb{N}_\tx{v}$, as defined in \eqref{eq:CC_joint}, \eqref{eq:CC_stage_based}, \eqref{eq:CC_node_based}. Lastly, \eqref{eq:OCP_initial_cond} defines the initial state, measured at the current time $t$.
In an open-loop setting, the above problem is solved once to yield the ego-vehicle control actions over the entire horizon $\uu_\tx{e}$. In the closed-loop setting, the problem is repeatedly solved with discrete time intervals $\Delta t$, applying the control action of the root node $\uu_0$.

% ---------------------------------------------------------------------------------------------
% ---------------------------------------------------------------------------------------------
% ---------------------------------------------------------------------------------------------
% ---------------------------------------------------------------------------------------------

\section{Simulation Study} \label{sec:results}
In the following section, we demonstrate the efficacy of our S-OCP approach in two case studies of interactive traffic scenarios. In the first, we investigate an unregulated road crossing where the ego-vehicle needs to negotiate crossing priority with a human driver. This study serves to demonstrate the implications of the different chance-constraint formulations and to provide empirical evidence for the presented tight outer-approximations,  displaying that the solution of the S-OCP indeed produces constraint violations according to $\varepsilon$. In the second, we investigate a highway lane merging scenario in dense traffic where the ego-vehicle needs to negotiate with another human driver to meet an upcoming highway exit. This study serves to demonstrate the efficacy of our approach in a more practical setting.

% % ---------------------------------------------------------------------------------------------
% % ---------------------------------------------------------------------------------------------
\subsubsection*{Reference Controllers}
To compare our proposed methods, we consider two alternative versions of problem \eqref{eq:S-OCP_reform}. The first, a \textit{Robust} alternative that does not utilize the estimated distribution \eqref{eq:p_hat}. Instead, the chance constraints are enforced for every state that has a strictly non-zero probability. This amounts to directly enforcing \eqref{eq:cor1_shortest_dist}, \eqref{eq:cor1_multipliers} for all nodes $\iota \in \mathbb{N}$. Similarly, all nodes in the objective receive equal weighting. The second, referred to as \textit{Approx}, utilizes the approach in \cite{wang2023interaction}, and approximates the indicator functions in the chance-constraints with a sigmoid function, i.e.,
\begin{equation*}
    \sum_{i=1}^{m} p_i \mathbf{1}_{(0,\infty)} \big ( g(\bxx_{i}) \big) < \sum_{i=1}^{m} p_i \sigma_{a,\alpha} \big ( g(\bxx_{i}) \big) < \varepsilon
\end{equation*}
where $\sigma_{a,\alpha}(x) = a/(1+\exp{(-\alpha x)})$. This constraint replaces \eqref{eq:OCP_cc_reformed}, but the problem is otherwise identical. The optimal outer approximation is indeed obtained by choosing $a=1+\exp{(0)}$, with $\alpha \to \infty$. To avoid numerical issues with large $\alpha$, we choose $\alpha = 3$.

\subsubsection*{Computational Resources} All experiments were conducted on a laptop equipped with an 12th Gen Intel(R) Core(TM) i7-12850HX CPU, and 32 GB of RAM. The different versions of \eqref{eq:S-OCP_reform} were formulated in CasADi \cite{casadi} and solved with IPOPT \cite{ipopt}. 
% % ---------------------------------------------------------------------------------------------
% % ---------------------------------------------------------------------------------------------

\subsection{Case 1: Unregulated Road Crossing}
\begin{figure}
    \centering
    \begin{subfigure}{0.49\linewidth}
    \includegraphics[width=1\linewidth,trim={2.2cm 0.4cm 2.2cm 0.4cm},clip]{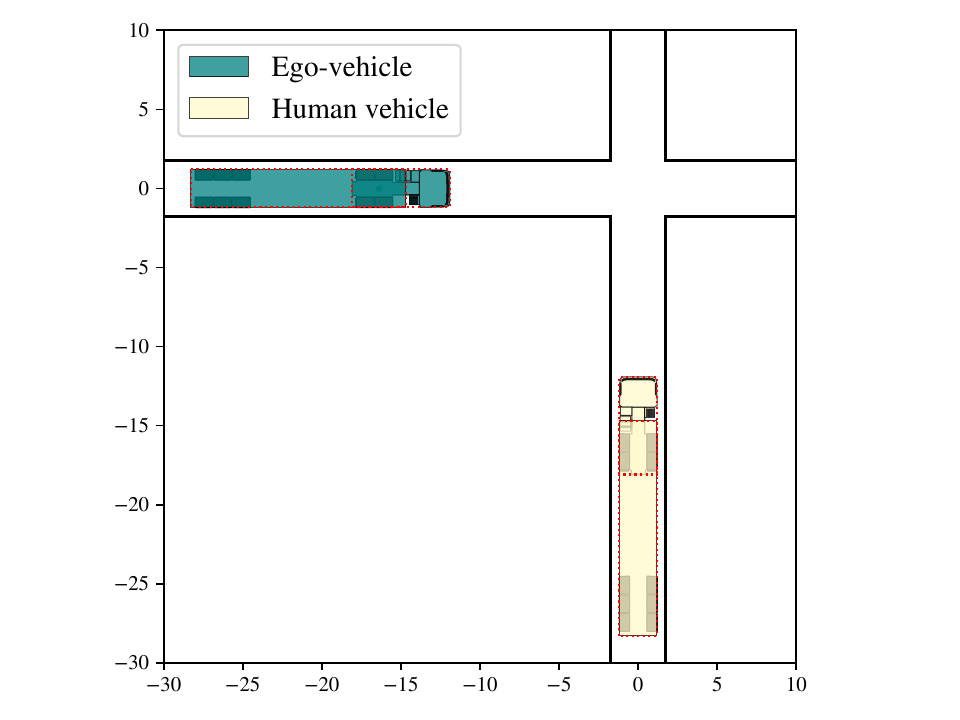}
    \label{fig:scene_two_way_intersection}
    \end{subfigure}
    \begin{subfigure}{0.49\linewidth}
    \includegraphics[width=1\linewidth,trim={0.35cm 0.8cm 0.2cm 0.35cm},clip]{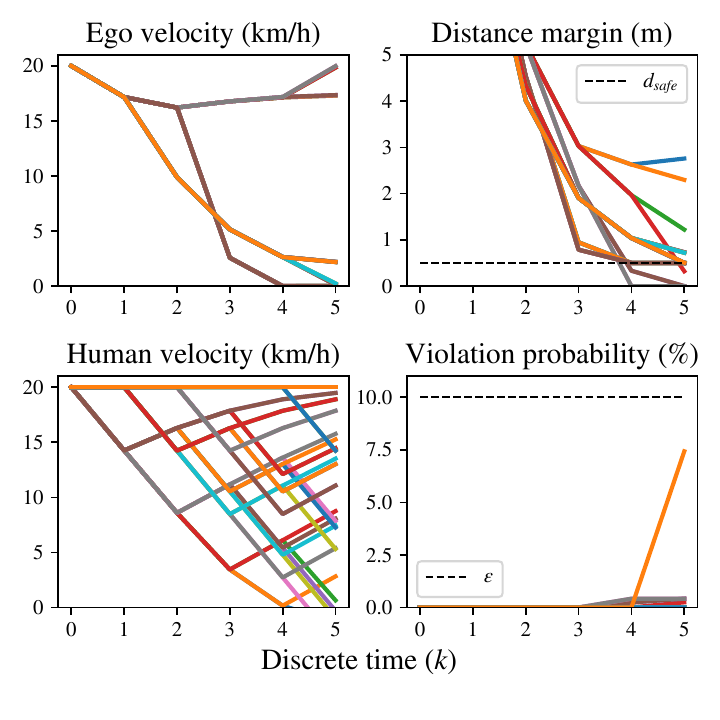}
    \label{fig:OCP_two_way_intersection}
    \end{subfigure}
    \caption{Initial conditions of road crossing case (left) and an S-OCP solution with $N=5$, using the joint chance constraint formulation (right). Each color indicates a path through the scenario tree, from the root node to a leaf node.}
    \label{fig:two_way_intersection}
\end{figure}

To demonstrate the theoretical guarantees of our approach we first investigate a simplified road crossing scenario without traffic rules such as, e.g., traffic lights and the right-hand-rule. Given that \eqref{eq:S-OCP_reform} is feasible at $t$, the solution of \eqref{eq:S-OCP_reform} is expected to have probabilistic guarantees that are asymptotically close to the theoretical formulations in Section \ref{sec:CCs_in_trees}. Further, given that we have correctly learned the distribution of human-driver decisions, simulations of the open-loop control strategy should exhibit a collision rate in accordance with the violation probability bound $\varepsilon$ and the choice of chance constraint version. To demonstrate these properties, we explore a safety-critical case in which two vehicles are imminently entering a road crossing, as displayed in Fig. \ref{fig:two_way_intersection}. The ego-vehicle adapts an open-loop control policy, originating from a solution of \eqref{eq:S-OCP_reform}, based on a ground truth model of the human driver.

\subsubsection{Simulation Details} The two vehicles are initialized \SI{15}{\meter} away from the crossing point, at a velocity of \SI{20}{\kilo\meter\per\hour}. The human driver picks from two possible decisions $\xi \in \{\tx{d_1},\tx{d}_2\}$, where $\tx{d}_1=\textit{braking}$ and $\tx{d}_2=\textit{tracking}$ is a reference velocity $v_\tx{ref}=\text{\SI{20}{\kilo\meter\per\hour}}$. The corresponding control laws $\kappa(\bxx,\xi)$ are based on the \textit{Intelligent Driver Model} (IDM) \cite{IDM}, tuned for an aggressive stop ($\tx{d}_1)$ or smooth tracking of a reference velocity ($\tx{d}_2$). As we aim to utilize the ground-truth distribution in the S-OCP, we utilize \eqref{eq:p_hat}, with features $\varphi(\bxx) = [p_{x,\tx{e}}/v_{\tx{e}},p_{y,\tx{h}}/v_{\tx{h}}]$ and weights $\theta_{\tx{d}_1}= [0.5,-0.5]^\top$, $\theta_{\tx{d}_2}= -\theta_{\tx{d}_1}$. This design yields $P_\xi(\bxx_0) = [0.5,0.5]^\top$.

\subsubsection{Controller Design}
\begin{table}[]
    \centering
    \caption{Shared Controller Parameters$^\mathrm{1}$.}
    \begin{tabular}{p{1.65cm}p{2.4cm}p{3.3cm}} \hline
         Parameter & Description & Values\\ \hline
         $\mathbf{Q}$ & State cost& $\tx{diag}(0,1,0.1,0,0)$\\
         $\mathbf{P}$& Terminal state cost & $\tx{diag}(0,1,0.1,180/\pi,180/\pi)$\\
         $\mathbf{R}$& Control cost& $\tx{diag}(1,180/\pi)$\\
         $\mathbf{R}_\Delta$& Control change cost & $\tx{diag}(0.1,0.1 \!\cdot \! 180/\pi)$\\
         $\xx_{\tx{e},\tx{ref}}$& State reference & [0, $y_\tx{ref}$, $v_\tx{ref}$, 0, 0] \\
         $[\underline{v}, \;\overline{v}]$& Velocity bounds &$[\mathrm{0},\mathrm{25}]$\SI{}{\kilo \meter \per \hour}\\
         $\big[\,\underline{\psi}_{1},\;\overline{\psi}_{1}\, \big]$& Tractor angle bounds &$[-\pi/8,\pi/8]$\SI{}{\radian}\\
         $\big [\underline{\psi}_{2},\;\overline{\psi}_{2} \big]$& Trailer angle bound &$[-\pi/8,\pi/8]$\SI{}{\radian}\\
         $[\underline{a},\overline{a}]$& Acceleration bound &$[-0.7\cdot9.8,0.05\cdot9.8]$\SI{}{\meter \per \second \squared}\\
         $[\underline{\delta},\;\overline{\delta}]$& Steering angle bound  &$[-\pi/8,\pi/8]$ \SI{}{\radian}\\
         $L_v$ & Total HVC Length & $L_1+L_2-(L_1/2-L_3)$ \\
         $L_1$ & HVC tractor length & \SI{6.18}{\meter}\\
         $L_2$ & HVC trailer length & \SI{13.60}{\meter}\\
         $L_3$ & Distance to hitch & \SI{1.39}{\meter}\\
         $l_w$ & HVC width & \SI{2.54}{\meter}\\
         $l_l$ & Lane width & \SI{3.75}{\meter} \\
         $d_\tx{safe}$ & Safety margin & $(l_l-l_w)/2$ \SI{}{\meter}\\
         $\varepsilon$& Violation Probability& 0.05\\\hline
    \end{tabular}
    \begin{tabular}{l}
            \footnotesize{$\!^\tx{1}$Values tuned for desirable performance across all cases and controllers.}
    \end{tabular}
    \label{tab:parameters_shared}
\end{table}

Each controller solves their respective version of \eqref{eq:S-OCP_reform} with parameters from Table \ref{tab:parameters_shared}. As we aim to utilize a correct model of the human driver, the scenario tree needs to describe all possible future states. Hence, we need to enumerate all possible decisions of the human driver and construct a tree without non-branching nodes, i.e $\mathbb{N}_\tx{br} = \NN  \setminus  \NN_N$. As the number of nodes increases exponentially with the length of the horizon, we opt for a short horizon $N=7$ with longer time steps $\Delta t=\text{\SI{0.7}{\second}}$. The ego-vehicle aims to track the lane center $y_\tx{ref}=0.0$ and a reference velocity $v_\tx{ref}=\text{\SI{20}{\kilo\meter\per\hour}}$. As intended, this yields an aggressive control strategy, where the ego-vehicle aims to cross before the human driver, introducing the possibility of collisions.

\subsubsection{Evaluation} The problem \eqref{eq:S-OCP_reform} is solved once for each controller version to obtain a control strategy $\uu_\tx{e}$ for each node in the tree. To evaluate efficacy, we perform simulations of the human driver, sampling the decision from $P_\xi(\bxx)$ over the prediction horizon $N$. Each simulation corresponds to a path through the tree from the root node to a leaf node. We define the \textit{Crossing Rate}, as the rate of sampled scenarios where the ego-vehicle crosses before the human driver without violating the constraints, and correspondingly define the \textit{Collision Rate} as the average number of sampled scenarios with one or more constraint violations. Further, we approximate the \textit{Expected Cost} by computing the average of \eqref{eq:OCP_objective} and we approximated the \textit{Expected Number of Constraint Violations}, henceforth abbreviated as ``\textit{ENCV}'', with the average number of constraint violations in each simulation. We evaluate seven different controller designs in total; Our proposed reformulation, referred to as \textit{Tight}, with the three different chance constraint formulations; The sigmoid approximation \textit{Approx}, with the three different chance constraints formulations, and the robust alternative \textit{Robust}.

\subsubsection{Results} A summary is provided in Table \ref{tab:res_two_way_intersection}. As a consequence of the cost function formulation, all controllers strive to keep the initial velocity, and cross before the human driver. However, crossing before the human requires risking collisions. Hence, the robust controller never crosses first and does not have any constraint violations. The node-based formulation, which does not propagate the probability from the root node, consequently utilizes a probability measure that does not reflect the likelihood of the sampled scenarios. In this case, the result is an overly conservative behavior. The stage-based and joint controller versions, with a correct estimation of the distribution of different scenarios, are able to relax constraints in sufficiently unlikely nodes and significantly improve the performance. The improvements can directly be seen from the evaluation of the cost function, but also indirectly via the rate at which the ego-vehicle crosses first. Simultaneously, the collision rate increases above zero but is kept below the specified $\varepsilon = 0.05$. Comparing \textit{Approx} with the \textit{Tight} reformulation we see that the approximations impose excessive limitations on the allowed probability of constraint violations, consequently limiting the performance. We can observe that Proposition \ref{prop:cc_reform} indeed can produce tight result by comparing the expected number of constraint violations (ENCV) for the Joint controller versions. In the approximation version, the expected number is close to zero, while in the tight version the ENCV approaches the specified $\varepsilon=0.05$.

\begin{table*}[ht]
    \vspace{4mm}
    \centering
    \caption{Averaged results for 10,000 simulations for the unregulated road crossing. Rates are presented as percentages (\%), costs are normalized.}
    \begin{tabular}{|c||c?c|c|c?c|c|c|}\hline
         Controller & \textit{Robust}& \textit{Approx} Node-based& \textit{Approx} Stage-based & \textit{Approx} Joint & \textit{Tight} Node-based & \textit{Tight} Stage-based & \textit{Tight} Joint\\ \hline \hline
         Crossing Rate & 0.00 & 0.00 & 46.24 & 47.05  & 0.00 & 49.96 & 47.57 \\\hline
         Collision Rate  & 0.00 & 0.00 & 3.72 & 0.00 & 0.00 & 4.94 & 1.72 \\\hline
         Expected Cost & 1.00 & 1.18 & 0.75 & 0.83 & 1.01 & 0.64 & 0.71\\\hline
         ENCV  & 0.00 & 0.00 & 1.13e-1 & 3.05e-5 & 0.0 & 1.62e-1 & 4.99e-2 \\\hline
    \end{tabular}
    \label{tab:res_two_way_intersection}
\end{table*}

% % ---------------------------------------------------------------------------------------------
% % ---------------------------------------------------------------------------------------------
\subsection{Case 2: Highway Lane Change}
\begin{figure}[t!]
    \vspace{1mm}
    \centering
    \begin{subfigure}{1 \textwidth}
    \includegraphics[width = 0.48 \textwidth,trim={0cm 3.5cm 0cm 4cm},clip]{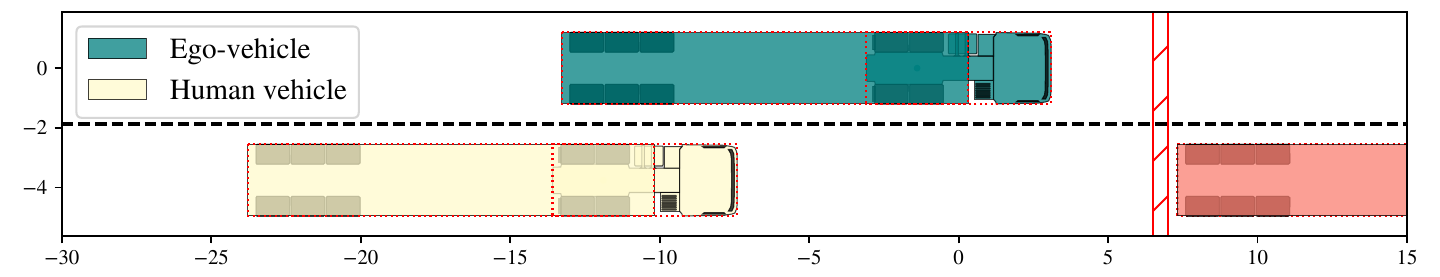}
    \end{subfigure}
    \begin{subfigure}{1 \textwidth}
    \includegraphics[width = 0.48 \textwidth,trim={0cm 3.5cm 0cm 4cm},clip]{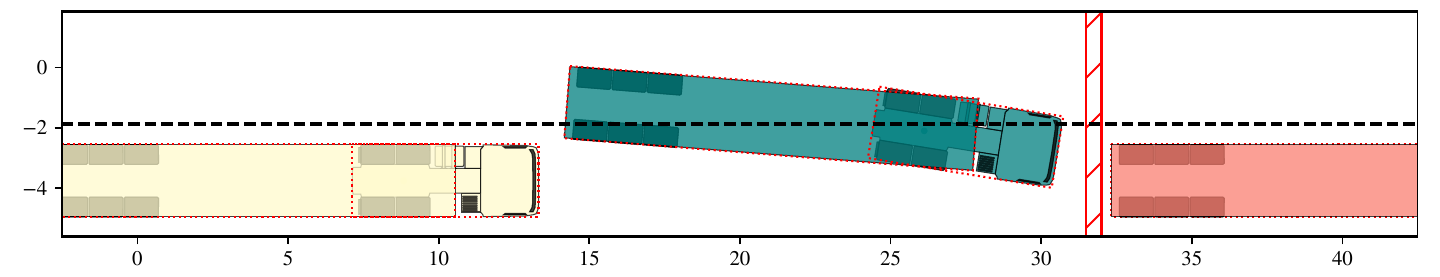}
    \end{subfigure}
    \begin{subfigure}{1 \textwidth}
    \includegraphics[width = 0.48 \textwidth,trim={0cm 3.5cm 0cm 4cm},clip]{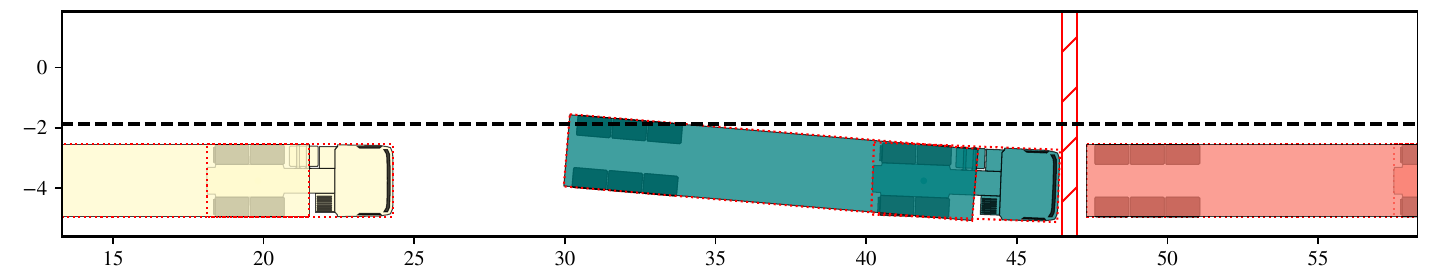}
    \end{subfigure}
    \caption{Example of a successful highway lane change. Each plot displays the ego-vehicle (teal), the interacting vehicle (beige) and the leading vehicle (red), at time $t$.}
    \label{fig:scene_highwayMerge}
\end{figure}
In this case, our aim is to investigate the closed-loop control performance in a practically relevant interactive scenario, with an approximated model of the human decision distribution. HVCs tend to conglomerate on highways, often forcing interactions in dense and slow-moving traffic. Here, we consider a case where the ego-vehicle is required to complete a lane change under a time constraint, e.g., to meet an upcoming highway exit. As displayed in Fig. \ref{fig:scene_highwayMerge}, the ego-vehicle is positioned in the left lane, wanting to move into a gap between two vehicles. As the leading vehicle $(\tx{l})$ is far ahead of the ego-vehicle, we assume that the ego-vehicle is unable to influence it's behavior. Hence, the ego-vehicle must encourage the adjacent human driver $(\tx{h})$ to decelerate, consequently increasing the available space and making the lane change feasible.

\subsubsection{Simulation Details}
All vehicles are initialized in their respective lane center at a velocity of \SI{20}{\kilo\meter\per\second}. The initial longitudinal position of the ego-vehicle is sampled as $p_{x,\tx{e}} \sim \mathcal{U}(-3,3)\text{\SI{}{\meter}}$ and, respectively, for the adjacent human driver $p_{x,\tx{h}} \sim \mathcal{U}(-10,-7)\text{\SI{}{\meter}}$. As the initial gap and the relation between the ego and the adjacent human-driver already vary, we initialize the leading vehicle deterministically at $p_{x,\tx{l}} = 7\text{\SI{}{\meter}}$. The adjacent human driver may similarly pick from two possible decisions $\xi \in \{\tx{d_1},\tx{d}_2\}$, where $\tx{d}_1=\textit{braking}$ and $\tx{d}_2=\textit{tracking}$ a reference velocity $v_\tx{ref}=\text{\SI{20}{\kilo\meter\per\hour}}$. The corresponding control laws $\kappa(\bxx,\xi)$ are based on the Predictive-IDM (P-IDM) \cite{brito2022learning}, tuned for a smooth deceleration ($\tx{d}_1)$ or smooth tracking of a reference velocity ($\tx{d}_2$). 

To model human decision-making, we introduce a stochastic version of the model in \cite{brito2022learning} as a parameterized distribution depending on the relative position of the ego and human vehicle,
\begin{equation*}
    \mathbb{P}(\xi = \tx{d}_1|\bxx) = \sigma((p_{x,\tx{e}}-p_{x,\tx{h}}-C_x)-(p_{y,\tx{e}}-p_{y,\tx{h}}-C_y)).
\end{equation*}
Further, $\mathbb{P}(\xi = \tx{d}_2|\bxx) = 1 - \mathbb{P}(\xi = \tx{d}_1|\bxx)$ and $\sigma(x) = 1/(1+\exp(-x)$ is a sigmoid function. The parameters $C_x$, $C_y$ are proportional to the HVCs length $L_v$ and the lane width $\ell_l$ respectively, as $C_x=(1-c)L_v$, $C_y=c \ell_l$, where $c \in [0,1]$ is a \textit{cooperation coefficient}, modulating the likelihood of interaction with cooperative versus non-cooperative human drivers. As our main aim is to highlight differences in the control design, we leave an accurate selection of $c$ for future work. In simulation, we observed $c=0.35$ to produce a high frequency of challenging interactive scenarios.

\subsubsection{Controller Design}
Each controller solves their respective version of \eqref{eq:S-OCP_reform} at each time step of the simulations, using parameters from Table \ref{tab:parameters_shared}. The ego-vehicle aims to track the adjacent lane's center $y_\tx{ref}=-\ell_l$ and a reference velocity $v_\tx{ref}=\text{\SI{20}{\kilo\meter\per\hour}}$. In this case study, we opt for a shorter time step $\Delta t = 0.3$ with a longer horizon $N=15$, which requires an approximate scenario tree. To balance computational complexity with modeling accuracy, we observed that choosing $|\mathbb{N}_\tx{br}| = 3$ with $\mathbb{N}_\tx{br} = \{0, \mathbb{N}_7\}$ provides a suitable tradeoff. Finally, the estimated distribution $\hat{P}(\bxx_\iota, \tx{d};{\boldsymbol{\theta}})$ is utilized in \eqref{eq:OCP_p_transition} to propagate the estimated probabilities through the considered scenario tree.

% Branching strategy

% Slack 

\subsubsection{Learning Design}
To learn the distribution of human decisions, we utilize \eqref{eq:p_hat} with features ${\varphi(\bxx) = [1, \xx_\tx{e}^\top-\xx_\tx{h}^\top]^\top}$. The optimization of the weights $\boldsymbol{\theta}$ is done offline, using Adam \cite{adam} with a learning rate $\tx{lr}=0.01$. Data were collected for 100 simulations of the highway lane change, utilizing the \textit{Robust} controller for the ego-vehicle. Data labeling was performed by observing changes in the human vehicles velocity. A strictly decreasing velocity was considered to correspond to $\tx{d_1}$, and $\tx{d_2}$ was considered otherwise.
\subsubsection{Evaluation} We ran 100 simulations of the driving scenario for each of the seven controller versions. Simulations were terminated if: The ego-vehicle reached the right lane's center; The vehicles' polytopes overlapped, or if a maximum simulation time of \SI{15}{\second} was reached. Similarly, we define the \textit{Success Rate}, \textit{Collision Rate}, and the \textit{Timeout Rate}, by averaging the outcome of all simulations. The \textit{Average Cost} is defined as the mean value of the cost function, computed over all states, and averaged across the 100 simulations.

\subsubsection{Results}

\begin{table*}[ht]
    \vspace{4mm}
    \centering
    \caption{Averaged results for 100 simulations of the highway lane change. Rates are presented as percentages (\%), and costs are normalized.}
    \begin{tabular}{|c||c?c|c|c?c|c|c|}\hline
         Controller & \textit{Robust}& \textit{Approx} Node-based& \textit{Approx} Stage-based & \textit{Approx} Joint & \textit{Tight} Node-based & \textit{Tight} Stage-based & \textit{Tight} Joint\\ \hline \hline
         Success Rate & 69 & 76 & 71 & 76 & 81 & 77 & 78 \\\hline
         Timeout Rate & 31 & 24 & 29 & 30  & 23 & 23 & 22 \\\hline
         Collision Rate & 0 & 0 & 0 & 0 & 0 & 0 & 0 \\\hline
         Average Cost  & 1.00 & 1.08 & 1.11 & 1.43 & 0.63 & 0.65 & 0.65 \\\hline
    \end{tabular}
    \label{tab:res_highwayMerge}
\end{table*}

A summary is provided in Table \ref{tab:res_highwayMerge}. All controllers are tracking the center of the adjacent lane's center and aim to complete the lane change as soon as possible. The feasibility and timing of the lane change depend on the predicted space between vehicles or, in the case of stochastic controllers, on the likelihood of the predicted space. Compared to the open-loop version in Case 1, the controllers seem to obtain a more similar performance in terms of success, timeout, and collision rate. Notably, all controllers avoided collisions, despite relying on a potentially incorrect distribution model derived by ML. In terms of success rate, the \textit{Tight} controllers outperformed the \textit{Approx}, which in turn surpassed the robust controller. Interestingly, the differences in performance between different chance constraints versions were minor. These observations seem to suggest that an accurate model of the stochastic environment was less critical for successful maneuver completion, compared to Case 1. The \textit{Approx} controllers obtained the highest average cost, while the lowest was obtained by the \textit{Tight} controllers.

\subsection{Conclusions and Future Work}
In a broad sense, the results suggest that incorporating chance-constrained stochastic optimal control has the potential to improve AV performance in interactive driving scenarios, compared to robust alternatives. The simulations indicate that the proposed learning-based closed-loop control strategy is capable of generating collision-free motion plans with high probability. Moreover, using tight constraint reformulations further improves performance compared to approximate methods, particularly in scenarios with narrow safety margins.
Although the presented learning-based S-OCP can provide promising performance, the probabilistic safety guarantees are lost when it is combined with ML.  Providing rigorous stochastic guarantees in the presence of machine learning components is hence a particularly interesting future work. We are futher interested in exploring more advanced branching strategies to improve the scenario tree approximation, and tuning the vehicle models using naturalistic driving data. Additionally, we plan to evaluate these approaches against high-fidelity simulators to further validate their effectiveness.

\section*{Acknowledgment}
The authors thank Morteza Haghir Chehreghani, Deepthi Pathare, Stefan Börjesson, Markus Gerdin, and Sten Elling Tingstad Jacobsen for insightful discussions about the research topic.

\appendices
\section{Proof of Proposition \ref{prop:joint_cc}} \label{A:proof_joint_cc}
We may express the probability of constraint violations over the transitions from a node $i\to \iota^+, \forall \iota^+ \in \tx{Ch}(\iota)$ as follows.
\begin{align*}
    &\mathbb{P}_{\zz_\iota} \Big [\bigcup_{\iota^+\in \tx{Ch}(\iota)} g(\xx_{\iota^+}) > 0 \, \big | \,\xx_\iota \Big ] = \\
    &\mathbb{E}_{\zz_\iota} \Big [\bigcup_{\iota^+\in \tx{Ch}(\iota)} \mathbf{1}_{(0,\infty)} \left (g(\xx_{\iota^+}) \right)\, \big | \,\xx_\iota \Big ] = \\
    &\sum_{\iota^+ \in \tx{Ch}(\iota)} p_{\iota \to \iota^+} \mathbf{1}_{(0,\infty)} \left (g(\xx_{\iota^+}) \right)
\end{align*}
Here,the last equivalence follows from the fact that the outcomes of $\zz_\iota$ are mutually exclusive. Further, we may apply the same idea recursively by considering a node $j$ and a set of nodes $\iota^+ \in \tx{Ch}(\tx{Ch}(\dots \tx{Ch}(j)))$ as follows.
\begin{align*}
    &\mathbb{P}_{\zz_j,\dots, \zz_\iota} \Big [\bigcup_{\iota^+ \in \tx{Ch}(\tx{Ch}(\dots \tx{Ch}(j)))} g(\xx_{\iota^+}) > 0 \, \big | \,\xx_j \Big ] = \\
    &\sum_{\iota^+ \in \tx{Ch}(\tx{Ch}(\dots \tx{Ch}(j)))} p_{j \to \iota^+} \mathbf{1}_{(0,\infty)} \left (g(\xx_{\iota^+}) \right)
\end{align*}
Similarly, the outcomes of $[\zz_j,\dots, \zz_\iota]$ are mutually exclusive for each $\iota^+$, and equivalence holds. Considering $j=0$, and a discrete time $k$ we have $\iota^+ \in \tx{Ch}(\tx{Ch}(\dots\tx{Ch}(0))) = \mathbb{N}_k$. Hence, we may provide the following outer approximation of the joint chance constraints,
\begin{align*}
    &\mathbb{P}_{\zz_0,\dots, \zz_N} \Big [\bigcup_{k=0}^{N-1}\bigcup_{\iota^+ \in \mathbb{N}_{k+1}} g(\xx_{\iota^+}) > 0 \, \big | \,\xx_0 \Big ] \leq \\
    &\sum_{k=0}^{N-1}\mathbb{P}_{\{\zz_0,\dots ,\zz_\iota, \forall \iota \in \mathbb{N}_k \}} \Big [\bigcup_{\iota^+ \in \mathbb{N}_{k+1}} g(\xx_{\iota^+}) > 0 \, \big | \,\xx_0 \Big ] = \\
    &\sum_{k=0}^{N-1}\mathbb{E}_{\{\zz_0,\dots ,\zz_\iota, \forall \iota \in \mathbb{N}_k \}} \Big [\bigcup_{\iota^+ \in \mathbb{N}_{k+1}} \mathbf{1}_{(0,\infty)} \left (g(\xx_{\iota^+}) \right) \, \big | \,\xx_0 \Big ] = \\
    &\sum_{k=0}^{N-1} \sum_{\iota^+ \in \mathbb{N}_{k+1}}p_{0 \to \iota^+} \mathbf{1}_{(0,\infty)} \left (g(\xx_{\iota^+}) \right)
\end{align*}
where Boole's inequality is applied in the first line. 

\section{Proof of Corollary \ref{cor:shortest_distance}} \label{A:proof_shortest_d}
With object representation \eqref{eq:V_rep}, \eqref{eq:dist} becomes a convex problem with affine constraints. Hence, Slater's condition \cite[Pages~226-227]{boyd2004convex}, and consequently strong duality holds. This implies that the optimal value of the dual objective \eqref{eq:thm_dist} is equivalent to the squared euclidean distance. 
% Following the proof of Theorem \ref{prop:dist2_dietz}, \eqref{eq:thm_dist} describes $-\tx{sd}(\mathbb{O}_{i,l},\mathbb{O}_{\hat{i},\hat{l}})$.
Hence, we may introduce an additional optimization variable $\gamma \in \RR$, as follows.
\begin{align} \nonumber
    & \min_{(l,\tilde{l}) \in \, \mathbf{I}_{N_{l,\tx{e}}} \times \,\mathbf{I}_{N_{\tilde{l},\tx{h}}}} \tx{dist}^2(\mathbb{O}_{\tx{e},l},\mathbb{O}_{\tx{h},\tilde{l}})  \geq d^2_\tx{safe} \Leftrightarrow \\ \nonumber
    & \max_{(l,\tilde{l}) \in \, \mathbf{I}_{N_{l,\tx{e}}} \times \,\mathbf{I}_{N_{\tilde{l},\tx{h}}}} -\tx{dist}^2(\mathbb{O}_{\tx{e},l},\mathbb{O}_{\tx{h},\tilde{l}})  \leq - d^2_\tx{safe} \Leftrightarrow \\ \nonumber
    &\min_{\gamma} \;\{\gamma:-\tx{dist}^2(\mathbb{O}_{i,l},\mathbb{O}_{\hat{i},\hat{l}})<\gamma, \; \forall (l,\hat{l})\} \leq - d_\tx{safe}^2 \Leftrightarrow\\ \nonumber
    &\exists \gamma \in \RR, \zeta \in\RR^{n \times N_c},\mu, \nu \in \RR^{N_c}: \gamma \leq -d^2_\tx{safe} \\ \label{eq:A_short_dist}
        & \begin{cases}
             \frac{1}{4}\zeta_{(l,\tilde{l})}^\top\zeta_{(l,\tilde{l})} + \mu_{(l,\tilde{l})} + \nu_{(l,\tilde{l})} \leq \gamma \\
             V_{\tx{e},l}(\xx_\tx{e})^\top\zeta_{(l,\tilde{l})} + \mu_{(l,\tilde{l})} \mathbf{1} \geq \mathbf{0}\\
             -V_{\tx{h},\tilde{l}}(\xx_\tx{h})^\top\zeta_{(l,\tilde{l})} + \nu_{(l,\tilde{l})} \mathbf{1} \geq \mathbf{0}
        \end{cases}\nquad
\end{align}

Here, line three applies the result in Proposition \ref{prop:dist2_dietz} and the \eqref{eq:A_short_dist} are repeated $\forall (l,\hat{l}) \!\!\in\! \mathbf{I}_{N_{l,\tx{e}}} \! \!\times \mathbf{I}_{N_{\hat{l},i}} \!\!\in \!\RR^{N_c}$. As 
the above problem remains convex, strong duality holds, and the optimal value $\gamma^*$ omits a tight approximation of the smallest squared distance, with a negative sign.\hfill $\square$

\section{Proof of Proposition \ref{prop:cc_reform}} \label{A:proof_cc_reform}
Consider parameters $p \in \RR^n$, $\varepsilon \in \RR$ and a function ${g: \RR^N \to \RR}$ with $\bxx_i \in \RR^{N}$ and $i \in \mathbb{N}_i = \{1, \dots,m\}$. We may then equivalently reformulate a constraint,
\begin{equation}\label{eq:prop_cc_general}
    \sum_{i=1}^{m} p_i \mathbf{1}_{(0,\infty)} \big ( g(\bxx_{i}) \big)  \leq \varepsilon
\end{equation}
by introducing variables $\epsilon \in \RR^m$ such that
\begin{align} \label{eq:prop_sum_reform}
    &\exists \epsilon \in \RR^m  : \; \sum_{i=1}^{m} \epsilon_i \leq \varepsilon, \; \epsilon_i \geq 0 \\\label{eq:prop_cc_split}
    & p_i \mathbf{1}_{(0,\infty)} \big ( g(\bxx_{i}) \big) \leq \epsilon_i, \; \forall i \in \mathbb{N}_i.
\end{align}
Note that the existence of any such $\epsilon$ ensures that \eqref{eq:prop_cc_general} is satisfied. Hence, we require a continuously differentiable and tight outer approximation of \eqref{eq:prop_cc_split}. To this end, we choose to utilize the hyperplane method \cite{fan2023efficient}. With variables $\lambda \in \RR^2$ and $\rho \in \RR$ we may express a half-space as
\begin{equation} \label{eq:prop_hyperplane}
    \lambda^T \begin{bmatrix}
        g(\bxx_i) \\
        p_i
    \end{bmatrix} < \rho.
\end{equation}
As displayed in Fig. \ref{fig:cc_reform}, the infeasible set of \eqref{eq:prop_cc_split} is a convex polytope with vertices $(0,\epsilon_{i}),(0,\varepsilon),(\infty,\varepsilon),$ and $(\infty,\epsilon_{i})$. Hence, we impose the following constraints
\begin{equation} \label{eq:prop_indicator_constraints}
    \lambda^\top \begin{bmatrix}
        0 & 0 & \infty & \infty \\
        \epsilon_i & \varepsilon & \varepsilon & \epsilon_i
    \end{bmatrix} \geq \rho \mathbf{1}^\top.
\end{equation}
\addtolength{\textheight}{-12cm}%
Indeed, for $g(\bxx_i)=0$ we wish that \eqref{eq:prop_hyperplane} imposes $p_i < e_i$. One may directly verify from \eqref{eq:prop_hyperplane} that this holds for a fixed intercept $\rho = e_i \lambda_2$ given $\lambda_2 \neq 0$. Similarly, we wish to avoid the trivial solution for $\lambda_1$ and enforce $\lambda_1 \neq 0$. In this setting, one may additionally verify that the constraints \eqref{eq:prop_indicator_constraints} then simplify as $\lambda_1\geq0$, $\lambda_2\geq0$. Expanding \eqref{eq:prop_hyperplane} and replacing \eqref{eq:prop_cc_split} yields
    \begin{align*}
        &\exists \epsilon \in \RR^{m},\lambda \in \RR^{2\times n}: \sum_{i=1}^m \epsilon_{i} \leq \varepsilon\\ \label{eq:prop1_ccs}
        &\begin{cases}
        \lambda_{1,i}g(\bxx_{i}) + \lambda_{2,{i}}\!\left(p_i - \epsilon_{i}\right) \! < 0 \\
        \lambda_{1,{i}} > 0,\;\lambda_{2,{i}} > 0\;, \epsilon_{i} \geq 0
        \end{cases} , \; \forall i \in \mathbb{N}_i
    \end{align*}
as desired. Note that the imposed hyperspace \eqref{eq:prop_hyperplane} implies that we may find some $\epsilon$, $\lambda$ such that $p_i$ may come arbitrarily close to $\epsilon_i$. Hence, the reformulation is tight.\hfill $\square$
\bibliographystyle{IEEEtran}
\bibliography{IEEEabrv,root.bib}

% ---------------------------------------------------------------------------------------------
% ---------------------------------------------------------------------------------------------
% ---------------------------------------------------------------------------------------------
% ---------------------------------------------------------------------------------------------

\end{document}

%% file: Figures/kinematic_model.tex
\newcommand\wheelW{1}
\newcommand\wheelH{0.4}

\newcommand\trailerAng{15.95}
\newcommand\steerAng{85}

\newcommand\textscale{0.7}

% Define CoG symbol
\tikzset{
  pics/CoG/.style args={#1,#2,#3}{
     code={
    % Input args : {radius, angle shift, color}
    \draw[#3,thick,fill=white] (0,0) circle[radius=#1];
    \tkzDefPoint(0,0){O}
    \tkzDefPoint({#1 * cos(pi/2+#2),#1 * sin(pi/2+#2)}){S1}
    \tkzDefPoint({#1 * cos(0.0+#2),#1 * sin(0.0+#2)}){S2}
    
    \tkzDefPoint({#1 * cos(pi/2+#2+pi),#1 * sin(pi/2+#2+pi)}){S3}
    \tkzDefPoint({#1 * cos(0.0+#2+pi),#1 * sin(0.0+#2+pi)}){S4}

    \tkzDrawSector[color=#3, thick,fill=#3](O,S2)(S1)
    \tkzDrawSector[color=#3, thick,fill=#3](O,S4)(S3)
     }
  }
}

\tikzset{
  pics/wheel/.style args={#1,#2,#3}{
     code={
    % Input args : {w, h, angle}
    \draw[thick,rounded corners = 2,fill = white,rotate around = {#3:(0,0)}] ({-#1/2,-#2/2}) rectangle (#1/2,#2/2);
     }
  }
}

\tikzset{
  ncbar/.style args={#1/#2/#3/#4}{ % expects: offset/text/position
    to path={
      % Draw the parallel shifted line
      ($(\tikztostart)!#1!90:(\tikztotarget)$)  -- 
      ($(\tikztotarget)!($(\tikztostart)!#1!90:(\tikztotarget)$)!90:(\tikztostart)$)
      % Place node with #2 text at midway, positioned #3 (above or below)
      node[midway, #3,
           anchor=#4] {\scalebox{\textscale}{#2}}
    }
  },
  ncbar/.default={0.5cm//hello//above//south east} % default offset 0.5cm, no text, above
}

\tikzset{
      orthLine/.style={
        to path=%
        ($(\tikztostart)!#1!90:(\tikztotarget)$)
        -- (\tikztostart)
      },
      orthLine/.default=0.5cm,
}

\tikzset{
  angle arc/.style args={#1/#2/#3/#4/#5/#6/#7}{ % A/C/radius/mid-pos/label
    insert path={
      let 
        \p1 = (#1),      % A
        \p2 = (#2),      % C
        \p3 = ($(\p1)!#4!(\p2)$), % Midpoint M (or any ratio)
        \n2 = {atan2(\y2-\y3,\x2-\x3)} % Angle from M to C
      in
      ([xshift=#3]\p3) arc[start angle=0, end angle=\n2, radius=#3]
        node[anchor=west, yshift=#6mm,xshift=#7mm] {\scalebox{\textscale}{#5}}

    }
  }
}

\tikzset{
  angle line/.style args={#1/#2/#3/#4}{ % A/C/radius/mid-pos/label
    insert path={
      let 
        \p1 = (#1),      % A
        \p2 = (#2),      % C
        \p3 = ($(\p1)!#4!(\p2)$), % Midpoint M (or any ratio)
        \n2 = {atan2(\y2-\y3,\x2-\x3)} % Angle from M to C
      in
      (\p3) -- ([xshift=#3]\p3)
    }
  }
}

\tikzset{
  line from angle/.style args={#1/#2/#3/#4/#5}{ % A/C/radius/mid-pos/angle
    insert path={
      let 
        \p1 = (#1),      % A
        \p2 = (#2),      % C
        \p3 = ($(\p1)!#4!(\p2)$), % Midpoint M (or any ratio)
        \n1 = {#3*cos(#5)},
        \n2 = {#3*sin(#5)},
        \p4 = (\n1,\n2),
        \n3 = {\x3 + \x4},
        \n4 = {\y3 + \y4},
        \p5 = (\n3,\n4),
        % \p5 = ($(\p3) + p5$)
      in
      (\p3) -- (\p5)
    }
  }
}

% arc from angle ={rearU1/frontU1/1cm/1/{\steerAng+180}/$\delta$/west/1/0}
\tikzset{
  arc from angle/.style args={#1/#2/#3/#4/#5/#6/#7/#8/#9}{ % A/C/radius/mid-pos/angle
    insert path={
      let 
        \p1 = (#1),      % A
        \p2 = (#2),      % C
        \p3 = ($(\p1)!#4!(\p2)$), % Midpoint M (or any ratio)
        \n1 = {#3*cos(#5)},
        \n2 = {#3*sin(#5)},
        \p4 = (\n1,\n2),
        \n3 = {\x3 + \x4},
        \n4 = {\y3 + \y4},
        \p5 = (\n3,\n4),
        \n5 = {atan2(\y1-\y2,\x1-\x2)}, % Angle from A to B
      in
        (\p5) arc[start angle=#5, end angle={\n5+180}, radius=#3]
        node[anchor=#7, yshift=#8mm,xshift=#9mm] {\scalebox{\textscale}{#6}}
    }
  }
}

\tikzset{
  arc between lines/.style args={#1/#2/#3/#4/#5/#6/#7/#8}{ % A/B (shared point)/C/radius/label
    insert path={
      let 
        \p1 = (#1),      % A
        \p2 = (#2),      % B
        \p3 = (#3),       % C
        \n1 = {atan2(\y1-\y2,\x1-\x2)}, % Angle from A to B
        \n2 = {atan2(\y3-\y2,\x3-\x2)}, % Angle from M to C
        \n4 = {\x2+#4*cos(\n1)},
        \n5 = {\y2+#4*sin(\n1)},
        \p4 = (\n4,\n5),
      in
      (\p4) arc[start angle=\n1, end angle=\n2, radius=#4]
      node[anchor=#6, yshift=#7mm,xshift=#8mm] {\scalebox{\textscale}{#5}}
    }
  }
}

\begin{tikzpicture}[
    roundnode/.style={circle, draw=blue, fill=blue,minimum size=1pt,,scale=0.3},
    box/.style={draw,fill=blue!10,thick,inner sep=3pt,minimum width=8em},
    coverbox/.style={fill=white,},yscale=1,xscale=1,xscale=1,
    ]
    
    % ------------------ Bicycle shape ------------------
    \coordinate (rearU1) at (-1.5,-1.5);
    \coordinate (frontU1) at (1.5,1.5);
    \coordinate (rearU2) at (-4,-1.5);
    \coordinate (frontU2) at (-0.5,-0.5);
    \coordinate (CoG) at (0,0);

    % ------------------ Coordinate system ------------------
    \draw[>=stealth,->,thick] (-5,-2.5) -- (3,-2.5) node[anchor=south west] {$X$};
    \draw[>=stealth,->, thick] (-5,-2.5) -- (-5,2.5) node[anchor=south west] {$Y$};
    \draw[gray,densely dotted] (CoG) to ([xshift=-5cm]CoG);
    \draw[gray,densely dotted] (CoG) to ([yshift=-2.5cm]CoG);

    % ------------------ Length Markers ------------------
    \draw[>=stealth, <->,ncbar={-1cm/$L_1$/below/north west}] (rearU1) to (frontU1);
    \draw[densely dotted,gray,semithick] (rearU1) to[orthLine=-1cm] (frontU1);
    \draw[densely dotted,gray,semithick] (frontU1) to[orthLine=1cm] (rearU1);
    
    \draw[>=stealth, <->,ncbar={0.6cm/$L_2$/above/south}] (rearU2) to (frontU2);
    \draw[densely dotted,gray,semithick] (rearU2) to[orthLine=0.6cm] (frontU2);
    \draw[densely dotted,gray,semithick] (frontU2) to[orthLine=-0.6cm] (rearU2);

    \draw[>=stealth, <->,ncbar={-0.4cm/$L_3$/below/north west}] (frontU2) to (CoG);
    \draw[densely dotted,gray,semithick] (frontU2) to[orthLine=-0.4cm] (CoG);
    \draw[densely dotted,gray,semithick] (CoG) to[orthLine=0.4cm] (frontU2);

    % ------------------ Angles ------------------
    % Velocity vector
    \coordinate (Vvector) at (0.5,1.1);
    \draw[>=stealth, ->, semithick] (CoG) to (Vvector) node [anchor=south] {\scalebox{\textscale}{$v$}};
    
    \begin{scope}[>={Stealth[scale=0.4]}, thick]
    % Tractor heading
    \draw[angle arc ={rearU1/frontU1/0.8cm/0.5/$\psi_1$/-1/1}, <->, semithick];
    \draw[angle line ={rearU1/frontU1/0.8cm/0.5}, semithick,densely dotted, gray];
    
    % Trailer heading
    \draw[angle arc ={rearU2/frontU2/1cm/0.3/$\psi_2$/-1/0}, <->, semithick];
    \draw[angle line ={rearU2/frontU2/1cm/0.3}, semithick,densely dotted, gray];

    % Lateral slip angle
    \draw[arc between lines ={frontU1/CoG/Vvector/0.8cm/$\beta$/west/1/0}, <->, semithick];

    % Steering angle
    \draw[arc from angle ={frontU1/rearU1/0.7cm/0.0/{\steerAng+180}/$\delta$/north west/-0.75/-0.75}, <->, semithick];
    
    \end{scope}

    \draw[line from angle ={rearU1/frontU1/0.7cm/1/{\steerAng+180}}, densely dotted, gray, semithick];
    % {frontU1/CoG/Vvector/0.8cm/$\beta$/west/1/0}
    % arc from angle

    % ------------------ CoG Markers ------------------
    \draw (CoG) pic{CoG={0.125,pi/4,black}};

    \node[anchor = south east,yshift=1mm,xshift=1mm] at (CoG) {\scalebox{\textscale}{$(p_x,p_y)$}};
    % \draw (-2.25,-1) pic{CoG={0.15,0.2782996590,black}};

    % ------------------ Bicycle shape ------------------
    \draw[black,thick] (rearU1) -- (frontU1);
    \node[roundnode,color = black, fill = black] at (frontU2) {};
    \draw[black,thick] (rearU2) -- (frontU2);

    % ------------------ Wheels ------------------
    \draw (rearU1) pic{wheel={\wheelW,\wheelH,45}};
    \draw (frontU1) pic{wheel={\wheelW,\wheelH,\steerAng}};
    \draw (rearU2) pic{wheel={\wheelW,\wheelH,\trailerAng}};
    
\end{tikzpicture}

%% file: Figures/scenario_tree.tex
\begin{tikzpicture}[
    roundnode/.style={circle, draw=blue,line width=0.75pt, fill=blue!40, minimum size=1pt,minimum size=1pt,scale=1.25},
    roundedbox/.style={draw,rounded corners,fill=gray!10,thick,inner sep=3pt,minimum width=8em},
    coverbox/.style={fill=white,},yscale=0.75,
    ]
% ------ k axis ------

\draw[->,thick] (-1,-4) -- (6.5,-4) node[right] {$k$};
\draw[-,dotted,gray!30,thick] (0,4.5) -- (0,-4);
\draw[-,dotted,gray!30,thick] (1,4.5) -- (1,-4);
\draw[-,dotted,gray!30,thick] (2,4.5) -- (2,-4);
\draw[-,dotted,gray!30,thick] (3,4.5) -- (3,-4);
\draw[-,dotted,gray!30,thick] (4,4.5) -- (4,-4);
\draw[-,dotted,gray!30,thick] (5,4.5) -- (5,-4);
\draw[-,dotted,gray!30,thick] (6,4.5) -- (6,-4);

\draw [-,thick] (0,-4cm-4pt) node[below] {\small $0$}-- (0,-4cm+4pt) ;
\foreach \x in {1,2,...,6} {%
    \draw [-] (\x,-4cm-2pt) -- (\x,-4cm+2pt) node[below,yshift = -4.5pt] {\small $\x$};
}
% ------ Branch Rectangles ------
% \node [roundedbox,draw,dashed,minimum width=0.5cm,minimum height=0.75cm] at (0,0) (B00) {};
% \node [roundedbox,draw,dashed,minimum width=2.5cm,minimum height=0.75cm] at (2,2) (B10) {};
% \node [roundedbox,draw,dashed,minimum width=2.5cm,minimum height=0.75cm] at (2,-2) (B11) {};
% \node [roundedbox,draw,dashed,minimum width=2.5cm,minimum height=0.75cm] at (5,3) (B20) {};
% \node [roundedbox,draw,dashed,minimum width=2.5cm,minimum height=0.75cm] at (5,1) (B21) {};
% \node [roundedbox,draw,dashed,minimum width=2.5cm,minimum height=0.75cm] at (5,-1) (B22) {};
% \node [roundedbox,draw,dashed,minimum width=2.5cm,minimum height=0.75cm] at (5,-3) (B23) {};

% ------------- Nodes -------------
% --- b = 0 ---
\node[roundnode,label = {[yshift=-0.45cm]\footnotesize 0},fill=red!25,draw=red,minimum size = 1mm] at (0,0) (000) {};
% --- b=1 ---
\node[roundnode,label = {[yshift=-0.45cm]\footnotesize 1}] at (1,2) (100) {};
\node[roundnode,label = {[yshift=-0.45cm]\footnotesize 3}] at (2,2) (101) {};
\node[roundnode,label = {[yshift=-0.45cm]\footnotesize 5},fill=red!25,draw=red] at (3,2) (102) {};

\node[roundnode,label = {[yshift=-0.45cm]\footnotesize 2}] at (1,-2) (110) {};
\node[roundnode,label = {[yshift=-0.45cm]\footnotesize 4}] at (2,-2) (111) {};
\node[roundnode,label = {[yshift=-0.45cm]\footnotesize 6},fill=red!25,draw=red] at (3,-2) (112) {};
% --- b=2 ---
\node[roundnode,label = {[yshift=-0.45cm]\footnotesize 7}] at (4,3) (200) {};
\node[roundnode,label = {[yshift=-0.45cm]\footnotesize 11}] at (5,3) (201) {};
\node[roundnode,label = {[yshift=-0.45cm]\footnotesize 15}] at (6,3) (202) {};

\node[roundnode,label = {[yshift=-0.45cm]\footnotesize 8}] at (4,1) (210) {};
\node[roundnode,label = {[yshift=-0.45cm]\footnotesize 12}] at (5,1) (211) {};
\node[roundnode,label = {[yshift=-0.45cm]\footnotesize 16}] at (6,1) (212) {};

\node[roundnode,label = {[yshift=-0.45cm]\footnotesize 9}] at (4,-1) (220) {};
\node[roundnode,label = {[yshift=-0.45cm]\footnotesize 13}] at (5,-1) (221) {};
\node[roundnode,label = {[yshift=-0.45cm]\footnotesize 17}] at (6,-1) (222) {};

\node[roundnode,label = {[yshift=-0.45cm]\footnotesize 10}] at (4,-3) (230) {};
\node[roundnode,label = {[yshift=-0.45cm]\footnotesize 14}] at (5,-3) (231) {};
\node[roundnode,label = {[yshift=-0.45cm]\footnotesize 18}] at (6,-3) (232) {};

% -------- Lines --------
% --- b=0-1 ---
\draw[->,gray] (000) -- (100) node[midway,above,xshift=-20pt,yshift = 2pt] {\color{red} \scriptsize $\mathbb{P}({\xi_0 \! = \! \tx{d}_1}|\bxx_0)$} ;
\draw[->,gray] (000) -- (110) node[midway,below,xshift=-20pt,yshift = -2pt] {\color{red} \scriptsize $\mathbb{P}({\xi_0 \! = \! \tx{d}_2}|\bxx_0)$};
% --- b=1 ---
\draw[->,gray] (100) -- (101) node[midway,above,yshift = 4pt,xshift = 0pt] {\color{blue} \scriptsize $\xi_1 \!= \!\tx{d}_1$};
\draw[->,gray] (101) -- (102) node[midway,above,yshift = 4pt,xshift = 0pt] {\color{blue} \scriptsize $\xi_3 \!= \!\tx{d}_1$};
\draw[->,gray] (110) -- (111) node[midway,above,yshift = 4pt,xshift = 0pt] {\color{blue} \scriptsize $\xi_2 \!= \! \tx{d}_2$};
\draw[->,gray] (111) -- (112) node[midway,above,yshift = 4pt,xshift = 0pt] {\color{blue} \scriptsize $\xi_4 \! = \! \tx{d}_2$};
% --- b=1-2 ---
\draw[->,gray] (102) -- (200) node[midway,above,xshift=-20pt,yshift = 3pt] {\color{red} \scriptsize $\mathbb{P}({\xi_5 \! = \! \tx{d}_1}|\bxx_5)$};
\draw[->,gray] (102) -- (210) node[midway,below,xshift=-20pt,yshift = -3pt] {\color{red} \scriptsize $\mathbb{P}({\xi_5 \! = \! \tx{d}_2}|\bxx_5)$};
\draw[->,gray] (112) -- (220) node[midway,above,xshift=-20pt,yshift = 3pt] {\color{red} \scriptsize $\mathbb{P}({\xi_6 \! = \! \tx{d}_1}|\bxx_6)$};
\draw[->,gray] (112) -- (230) node[midway,below,xshift=-20pt,yshift = -3pt] {\color{red} \scriptsize $\mathbb{P}({\xi_6 \! = \! \tx{d}_2}|\bxx_6)$};
% --- b=2 ---
\draw[->,gray] (200) -- (201) node[midway,above,yshift = 4pt,xshift = 0pt] {\color{blue} \scriptsize $\xi_7 \!= \!\tx{d}_1$};
\draw[->,gray] (201) -- (202) node[midway,above,yshift = 4pt,xshift = 2pt] {\color{blue} \scriptsize $\xi_{11} \!= \!\tx{d}_1$};
\draw[->,gray] (210) -- (211) node[midway,above,yshift = 4pt,xshift = 0pt] {\color{blue} \scriptsize $\xi_8 \!= \!\tx{d}_2$};
\draw[->,gray] (211) -- (212) node[midway,above,yshift = 4pt,xshift = 2pt] {\color{blue} \scriptsize $\xi_{12} \!= \!\tx{d}_2$};

\draw[->,gray] (220) -- (221) node[midway,above,yshift = 4pt,xshift = 0pt] {\color{blue} \scriptsize $\xi_{9} \!= \!\tx{d}_1$};
\draw[->,gray] (221) -- (222) node[midway,above,yshift = 4pt,xshift = 2pt] {\color{blue} \scriptsize $\xi_{13} \!= \!\tx{d}_1$};
\draw[->,gray] (230) -- (231) node[midway,above,yshift = 4pt,xshift = 0pt] {\color{blue} \scriptsize $\xi_{10} \!= \!\tx{d}_2$};
\draw[->,gray] (231) -- (232) node[midway,above,yshift = 4pt,xshift = 2pt] {\color{blue} \scriptsize $\xi_{14} \!= \!\tx{d}_2$};

% --- Adding text ---

% % --- Cover boxes ---
% \node [coverbox,minimum width = 1cm,minimum height = 1.1cm] at (6.25,3) (cB20) {};
% \node [coverbox,minimum width = 1cm,minimum height = 1.1cm] at (6.25,1) (cB20) {};
% \node [coverbox,minimum width = 1cm,minimum height = 1.1cm] at (6.25,-1) (cB20) {};
% \node [coverbox,minimum width = 1cm,minimum height = 1.1cm] at (6.25,-3) (cB20) {};

\end{tikzpicture}

%% file: Figures/cc_reform.tex
\begin{tikzpicture}[
    roundnode/.style={circle, draw=blue, fill=blue,minimum size=1pt,,scale=0.4},
    box/.style={draw,fill=blue!10,thick,inner sep=3pt,minimum width=8em},
    coverbox/.style={fill=white,},yscale=0.3,
    ]

% ------ Coordinate System ------
\draw[>=stealth,->] (-3,0) -- (4,0) node[right] {$g(\bxx)$};
\draw[>=stealth,->] (0,0) -- (0,10) node[right] { };

% Unsafe set box
% % Fill
\node [box,minimum width = 3cm,minimum height = 1.4cm,draw=white,color = red!25] at (1.5,7) (iB) {};
\node [box,minimum width = 1cm,minimum height = 1.4cm,draw=white,color = red!20] at (3.5,7) (iB2) {};

\foreach \y in {5,6,...,9} {%
    \draw [-,dashed,color = red!50] (3,\y) -- (4,\y);
}

% % Outline
\draw[color = red!80] (0,4.6) node[left,yshift = 4pt] {$(0, \epsilon_i)$}-- (4,4.6) node[right,yshift = 4pt] {$(\infty, \epsilon_i)$};
\draw[color = red!80] (0,9.4) node[left,yshift = 4pt] {$(0, \varepsilon)$} -- (4,9.4);
\draw[color = red!80] (0,4.6) -- (0,9.4);
% \draw[color = red!80] (4,4.6) -- (4,9.4);

% % Coordinates
\node[roundnode,color = red,fill=white,thick] at (0,4.6) (01) {};
\node[roundnode,color = red,fill=white,thick] at (0,9.4) (02) {};
\node[roundnode,color = red,fill=white,thick] at (4,4.6) (03) {};

% Example hyperplane
\node[roundnode,color = blue,fill=white,thick] at (0,3) (H1) {};

\draw[color = blue!60,thick] (H1) node[left,yshift = -6pt] {$(g(\bxx_i),p_i)$} -- (3,1);
\draw[color = blue!60,thick] (H1) -- (-3,5);
\draw[color = blue!60,thick,dashed] (3,1) -- (4,0.33);

% ------ CVar approximation ------
% Last section, ploted below indicator func
% \draw[thick,color = red] (-0.5,2.25)  node[above,xshift = -1.5cm, yshift = 0.2cm] {$\hat{P}_\tx{m}(\mathbf{s};\theta) \left(1+ \alpha^* g(\mathbf{s^+}) \right)$} -- (3,7.5);
% \draw[thick,color = red,dashed] (3,7.5)  -- (4,9);

% \draw[color = black,dashed] (0,6) -- (4,6);
% \draw[color = black] (-2pt,6) node[left] {\small $(0,\epsilon_i)$} -- (2pt,6);

% ------ Indicator function ------
% \node[roundnode,label = below:$0$] at (0,0) (00) {};
% \node[roundnode,fill=white,thick] at (0,3) (01) {};
% \draw[thick,color = blue] (01) -- (3,3);
% \draw[thick,color = blue] (00) -- (-3.5,0);
% \draw[thick,color = blue,dashed] (3,3) node[above] {$\hat{P}_\tx{m}(\mathbf{s};\theta) \mathbbm{1}_{(\!0,\infty\!)}\!(g(\bxx))$}-- (4,3);
% \draw[thick,color = blue,dashed] (-3.5,0)--(-4,0);

% ------ CVar approximation ------
% - First section, displayed above indicator function
% \draw[thick,color = red] (-2,0) -- (-0.5,2.25);
% \draw[color = black] (-2,5pt) -- (-2,-5pt) node[below] {\small $g(\mathbf{s^+})$};

% \draw[thick,color = red] (-2,0) -- (-2.25,0);
% \draw[thick,color = red,dashed] (-2.25,0) -- (-3.9,0);
% \draw[thick,color = red,dashed] (-2,0) -- (-2.55,-1);

\end{tikzpicture}